\documentclass[10pt,twocolumn,letterpaper]{article}

\usepackage{cvpr} 
\usepackage[accsupp]{axessibility}

\usepackage[dvipsnames]{xcolor}

\definecolor{cvprblue}{rgb}{0.21,0.49,0.74}
\usepackage[pagebackref,breaklinks,colorlinks,citecolor=cvprblue]{hyperref}
\usepackage{amsmath,amsfonts}
\usepackage{algorithmic}
\usepackage{algorithm}
\usepackage{array}
\usepackage{multirow}
\usepackage{booktabs}
\usepackage{textcomp}
\usepackage{url}
\usepackage{verbatim}
\usepackage{graphicx}
\usepackage{cite}
\usepackage{epsfig}
\usepackage{amssymb}
\usepackage{float}
\usepackage{color}
\usepackage{tabularx}
\usepackage{longtable}
\usepackage{tabu}
\usepackage{bbding}
\usepackage{stfloats}
\usepackage{bbm}
\usepackage{color}
\usepackage{xcolor}
\usepackage{colortbl}
\definecolor{tablegray}{gray}{.9}
\usepackage{threeparttable}
\usepackage{longtable}
\usepackage{bm}

\makeatletter
\DeclareRobustCommand\onedot{\futurelet\@let@token\@onedot}
\def\@onedot{\ifx\@let@token.\else.\null\fi\xspace}
\def\eg{\emph{e.g}\onedot} 
\def\ie{\emph{i.e}\onedot}

\def\etal{\emph{et al}\onedot}
\newcommand*{\rom}[1]{\expandafter\@slowromancap\romannumeral #1@}
\makeatother

\def\paperID{10691} 
\def\confName{CVPR}
\def\confYear{2024}

\title{Perceptual Assessment and Optimization of HDR Image Rendering}

\author{Peibei Cao$^{1}$, Rafa{\l} K. Mantiuk$^{2}$, and Kede Ma$^{1,3}\thanks{Corresponding author.}$\\
$^1$ Department of Computer Science,  City University of Hong Kong \\
$^2$ Department of Computer Science and Technology,  University of Cambridge\\
$^3$ Shenzhen Research Institute,  City University of Hong Kong\\
\texttt{peibeicao2-c@my.cityu.edu.hk, kede.ma@cityu.edu.hk,  rkm38@cam.ac.uk}\\
\url{https://github.com/cpb68/HDRQA/}
}

\begin{document}
\maketitle

\begin{abstract}
High dynamic range (HDR) rendering has the ability to faithfully reproduce the wide luminance ranges in natural scenes, but how to accurately assess the rendering quality is relatively underexplored. Existing quality models are mostly designed for low dynamic range (LDR) images, and do not align well with human perception of HDR image quality. To fill this gap, we propose a family of HDR quality metrics, in which the key step is employing a simple inverse display model to decompose an HDR image into a stack of LDR images with varying exposures. Subsequently, these decomposed images are assessed through well-established LDR quality metrics. Our HDR quality models present three distinct benefits. First, they directly inherit the recent advancements of LDR quality metrics. Second, they do not rely on human perceptual data of HDR image quality for re-calibration. Third, they facilitate the alignment and prioritization of specific luminance ranges for more accurate and detailed quality assessment. Experimental results show that our HDR quality metrics consistently outperform existing models in terms of quality assessment on four HDR image quality datasets and perceptual optimization of HDR novel view synthesis. 
\end{abstract}

\section{Introduction}
High dynamic range (HDR) images aim to faithfully capture the large luminance variations of natural scenes that low dynamic range (LDR) images are not capable of~\cite{hoefflinger2007high}. In the past few years, numerous HDR imaging and display devices have been developed and commercialized in response to the escalating demand for HDR images in various fields, including photography, gaming, film, and virtual reality. Consequently, HDR image quality assessment (IQA) has become a practically demanding technique for monitoring, ensuring, and optimizing the perceptual quality of HDR images during imaging, compression, communication, and rendering. 

At present, HDR quality metrics are largely lacking, which is likely due to the prevailing assumption that most LDR quality models, such as the peak signal-to-noise ratio (PSNR) and the structural similarity (SSIM) index~\cite{wang2004image},
are readily applicable to HDR images. It was not until recently that researchers realized their poor account for human perception of HDR image quality~\cite{eilertsen2021cheat,Hanji2022}. Mantiuk~\etal~\cite{mantiuk2005predicting} made initial attempts by extending the classic visual difference predictor (VDP)~\cite{Daly1992} to HDR-VDP,
which was subsequently improved from various psychophysical and physiological perspectives~\cite{Mantiuk2011hdrvdp2,narwaria2015hdr,mantiuk2023hdr}.
Although the HDR-VDP family embodies many aspects of the early visual system, they contain complex and non-differentiable modules, which may hinder their application scope, especially when adopted as loss functions in perceptual optimization.

Other initiatives have focused on transforming linear luminances into a perceptually more uniform space as a way of improving the applicability of LDR quality metrics. Representative transformations include the logarithmic function~\cite{xu2005high}, the perceptually uniform (PU) encoding curve~\cite{aydin2008extending} and its derivative, the PU21 encoding~\cite{azimi2021pu21}, and the perceptual quantizer~\cite{miller2013perceptual}. 
The issue with perceptually uniform transformations lies in their tendency to either map luminance values that surpass the LDR image range (\eg, PU21 assigns high luminances to values above $255$), or to compress the values to a range of $[0,1]$ (\eg, the perceptual quantizer), which alters the image contrast. In the former case, image quality models incapable of handling values beyond the maximum pixel value ($255$ or $1$) will fail to capture distortions in bright regions. In the latter case, the compressed contrast will cause unanticipated changes in the metric predictions.

\begin{figure}[!t]
\centering
  \includegraphics[width=0.47\textwidth]{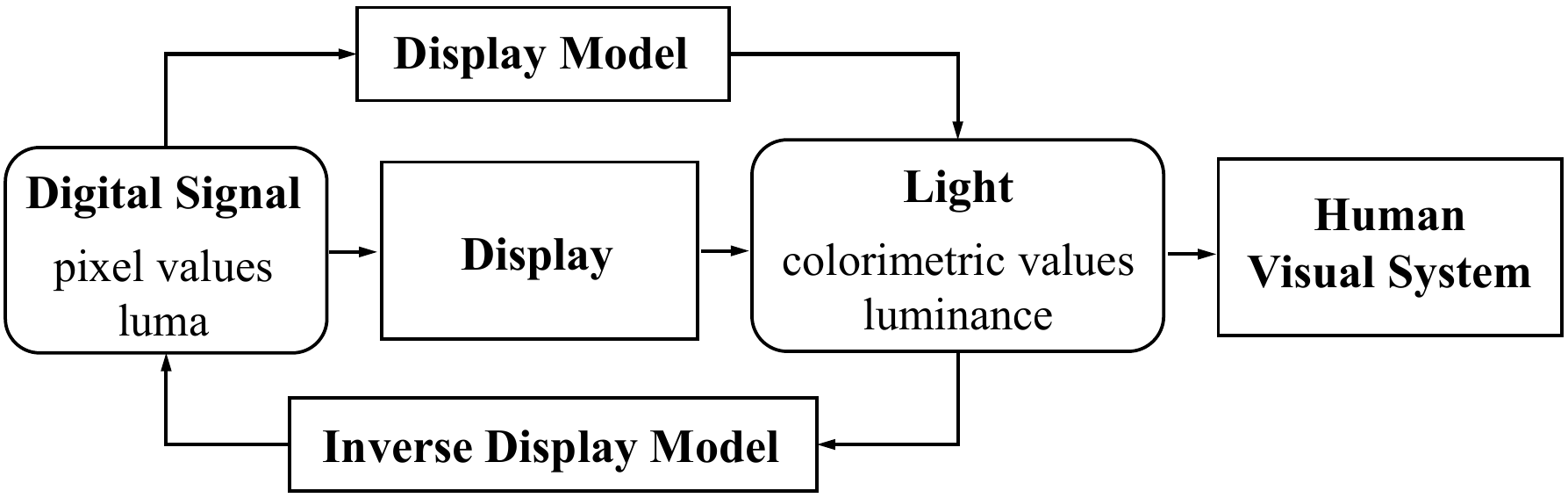}
  \caption{Forward display model simulates the process of converting digital pixel values into physical light in luminances on display. An inverse display model provides the inverse mapping.}
  \label{fig:IDM}
  \vspace{-0.8em}
\end{figure}

Inspired by~\cite{Munkberg06}, we propose a family of full-reference HDR quality metrics, which rely on a simple inverse display model~\cite{mantiuk2009visualizing} to transform an (uncalibrated) HDR image to a stack of LDR images with varying exposures, amenable to LDR-IQA. 
Our HDR quality metrics offer several key advantages. First, they enjoy the latest developments of and reduce gracefully to LDR quality metrics, with the adoption of a complementary display model (see Fig.~\ref{fig:IDM}). Second, they do not need human perceptual data of HDR image quality for re-calibration. Third, they allow for the weighting of specific luminance ranges to highlight their contributions during quality assessment and perceptual optimization. Fourth, they enable the efficient mitigation of possible luminance shifts between the reference and test HDR images for more accurate quality assessment~\cite{eilertsen2021cheat,Hanji2022}. Experimental results on four human-rated HDR-IQA datasets confirm the superior performance of our metrics, compared to existing models including the HDR-VDP family. We further demonstrate the promise of our HDR quality metrics as the perceptual optimization objectives in HDR novel view synthesis~\cite{huang2022HDRnerf}. Importantly, we observe a significant improvement in visual quality for over-exposed regions, which is corroborated by subjective user studies and objective quality estimates.

\section{Related Work}
\label{sec: related work}
In this section, we review two bodies of studies that are related to ours, HDR-IQA and HDR novel view synthesis.

\subsection{HDR Quality Metrics}

\noindent\textbf{Model-based methods} rely on computational models that emulate the physiological responses of neurons in the human visual system, particularly those in the primary visual cortex. HDR-VDP~\cite{mantiuk2005predicting} is an excellent example that takes into account aspects of nonlinear photoreceptor response to light, contrast sensitivity, and local adaptation. Similar to VDP~\cite{Daly1992}, HDR-VDP predicts visible difference maps without supplying a numerical quality score. HDR-VDP-2~\cite{Mantiuk2011hdrvdp2} improves upon HDR-VDP with a revised model of the early visual system. The metric was trained on two LDR-IQA datasets (\ie, LIVE~\cite{Sheikh2006Statistical} and TID2008~\cite{Ponomarenko2009Metrics}), and was later retrained on two additional HDR-IQA datasets: Narwaria2013~\cite{narwaria2013tone} and Narwaria2014~\cite{Narwaria2014impact}, leading to HDR-VDP-2.2~\cite{narwaria2015hdr}. More recently, HDR-VDP-3~\cite{mantiuk2023hdr} was developed by simulating the impact of aging on the visual system~\cite{Mantiuk2018age}, modeling the effect of adaptation to local luminances~\cite{Vangorp2015Model}, and re-calibrating the metric on the largest HDR-IQA dataset, UPIQ~\cite{mikhailiuk2021consolidated}. Other representative model-based methods include the HDR video quality measure (HDR-VQM)~\cite{HDR-VQM} and the normalized Laplacian pyramid distance (NLPD)~\cite{laparra2017perceptually}. Like  HDR-VDP, HDR-VQM follows an error visibility paradigm with the PU encoding as the front-end processing, while NLPD incorporates divisive normalization as a form of local gain control~\cite{carandini2012normalization}.

\begin{figure*}
\centering
  \includegraphics[width=\textwidth]{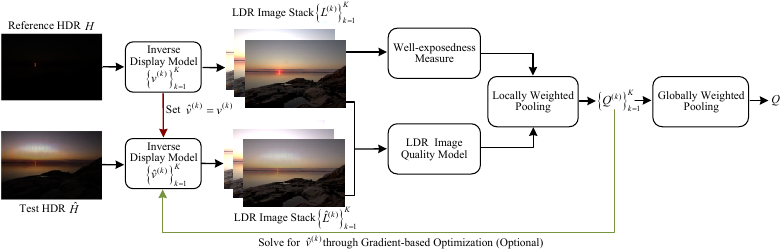}
  \caption{System diagram of the proposed family of HDR-IQA metrics. The default red arrow can be replaced by the optional green arrow, whose goal is to compensate for the possible luminance shifts between the reference and test HDR images, similar to the camera response function correction in~\cite{eilertsen2021cheat,Hanji2022}.}
  \label{fig:loss}
  \vspace{-0.8em}
\end{figure*}

\noindent\textbf{Encoding-based methods} transform linear luminances into a perceptually more uniform space for subsequent processing. Xu~\etal~\cite{xu2005high} approximated the luminance response curve as a logarithmic function. The PU encoding~\cite{aydin2008extending} was derived from the contrast sensitivity function (CSF) in~\cite{Daly1992}, which was optimized to approximate the gamma-encoding in the range from $0.1$ to $80$ $\rm cd/m^{2}$. Similarly, the perceptual quantizer~\cite{miller2013perceptual} was based on the Barten’s CSF~\cite{Barten03}, and was standardized in the ITU-R Recommendation BT.$2100$. As an improved version, PU21 encoding~\cite{azimi2021pu21} relies on a latest CSF~\cite{mantiuk2020practical}, which predicts contrast thresholds at luminance levels between $0.0002$ and $10,000$ $\rm cd/m^{2}$.  
Nevertheless, the PU encoding is designed for the luminance channel only, and is less applicable to chromatic channels.

Our family of HDR quality metrics falls naturally in the category of encoding-based methods. Inspired by Munkberg~\etal~\cite{Munkberg06}, we ``encode" an HDR image into a multi-exposure LDR image stack for reliable LDR-IQA.

\subsection{Novel View Synthesis}
Novel view synthesis, a typical application of image-based rendering, involves generating images from novel viewpoints given a set of input views~\cite{shum2000review}. The view synthesis can be performed directly in the pixel domain when the input images are densely sampled~\cite{Gortler1996Lumigraph, Levoy1996light}. It is more common and economic to capture inputs from a wider range of sparse locations, which will be processed through a ``proxy" geometry using either a heuristic~\cite{Buehler2001unstructured} or learned blending function~\cite{Hedman2018Deep, Riegler2020Free, Riegler2021Stable}.

Of particular interest in this line of research is NeRF, which represents a scene with a neural radiance field~\cite{Mildenhall2021NeRF,barron2021mip, boss2021nerd, chen2022hallucinated, li2021neural, ma2022deblur, martin2021nerf, yu2021pixelnerf}. Mildenhall~\etal~\cite{Mildenhall2021NeRF} demonstrated that neural implicit representations yield superior results in view synthesis compared to traditional explicit representations such as point clouds, voxels, and octrees. Various aspects of NeRF have been improved, including rendering quality and capability~\cite{MipNeRF360Barron2022}, training and rendering efficiency~\cite{zhenxing2022switch,barron2021mip}, robustness to varying illumination~\cite{martin2021nerf} and deformable objects~\cite{park2021nerfies}, compositionality~\cite{niemeyer2021giraffe}, editability~\cite{liu2021editing}, and generalization to novel scenes~\cite{chen2021mvsnerf}.

Recently, NeRF has been extended to work with HDR image data~\cite{mildenhall2022nerf,huang2022HDRnerf}.  Mildenhall~\etal~\cite{mildenhall2022nerf} trained  Mip-NeRF~\cite{barron2021mip} using linear noisy RAW images. Huang~\etal~\cite{huang2022HDRnerf} modeled the physical imaging process with two implicit functions: a radiance field and a tone mapper, which are jointly optimized taking multiple LDR images with different exposures as inputs. In this paper, we simplify Huang's method~\cite{huang2022HDRnerf} by stripping off the tone mapper and directly optimizing the RAW radiance field guided by the proposed HDR quality metrics.

\section{Proposed HDR Quality Metrics}
\label{sec:Approach}
In this section, we propose to transform the problem of HDR-IQA into LDR-IQA, with the help of a simple inverse display model~\cite{mantiuk2009visualizing}. Fig.~\ref{fig:loss} shows the system diagram of the proposed family of HDR-IQA metrics.

\subsection{Inverse Display Model}\label{sec:HDR}
A forward display model simulates how the display transforms digital pixel values to physical units of light, while the opposite mapping from physical units to digital values, is referred to as an inverse display model, as illustrated in Fig.~\ref{fig:IDM}. Here, we resort to an inverse gain-offset-gamma display model~\cite{mantiuk2009visualizing}:
\begin{equation}
\label{eq:IDM}
L^{(k)}= \left(\left[\frac{H\cdot v^{(k)}-b}{1-b}\right]_0^1\right)^{\frac{1}{\gamma}},  \quad 1\le k \le K,
\end{equation}
where $v^{(k)}$ is the $k$-th exposure value, determining the position of the dynamic range window to be mapped to the available luminance range of the display. 
We assume a fixed display device with the minimum and maximum luminances of $I_\mathrm{min} = 1$\,$\mathrm{cd}/\mathrm{m}^2$ and $I_\mathrm{max} = 200$ $\mathrm{cd}/\mathrm{m}^2$, respectively. These are typical specifications of consumer-grade displays of standard dynamic ranges, resulting in the window size $w=\log_2(200/1)=7.64$ in the logarithmic scale. $H$ denotes the reference HDR image, and $L^{(k)}$ represents the $k$-th LDR image. $b$ indicates the black-level factor, accounting for the limited contrast of the display due to the light leakage and the ambient light reflections from the display.
$[\cdot]_0^1$ denotes the clamping function with the output range $[0,1]$. $(\cdot)^{1/\gamma}$ represents the gamma correction. We follow the default configurations in~\cite{mantiuk2009visualizing}, and set $b=1/128$ and $\gamma=2.2$. Eq.~\eqref{eq:IDM} is independently applied to the three color channels.

It is noteworthy that we intentionally avoid employing state-of-the-art tone mapping operators (TMOs) for the HDR-to-LDR conversion. This is because they are essentially dynamic range compressors, leading to the unavoidable loss of information and the emergence of algorithm-dependent artifacts. In contrast, the adopted inverse display model incurs minimal contrast distortions by mapping a portion of the luminance range to that of the LDR display. Moreover, it acts as a local dynamic range magnifier, expanding a specific luminance range for a more detailed examination.

\begin{figure}[!t]
\centering
\subfloat[]{\includegraphics[width=1\linewidth]{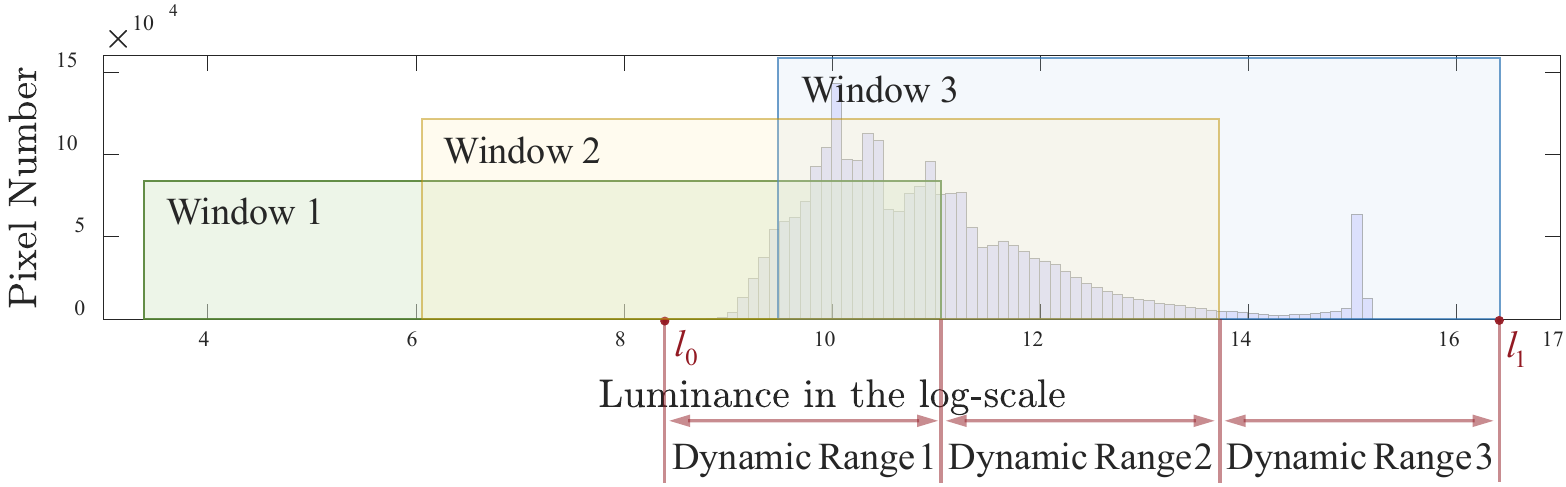}}\hskip.3em
\\
\subfloat[]{\includegraphics[width=0.32\linewidth]{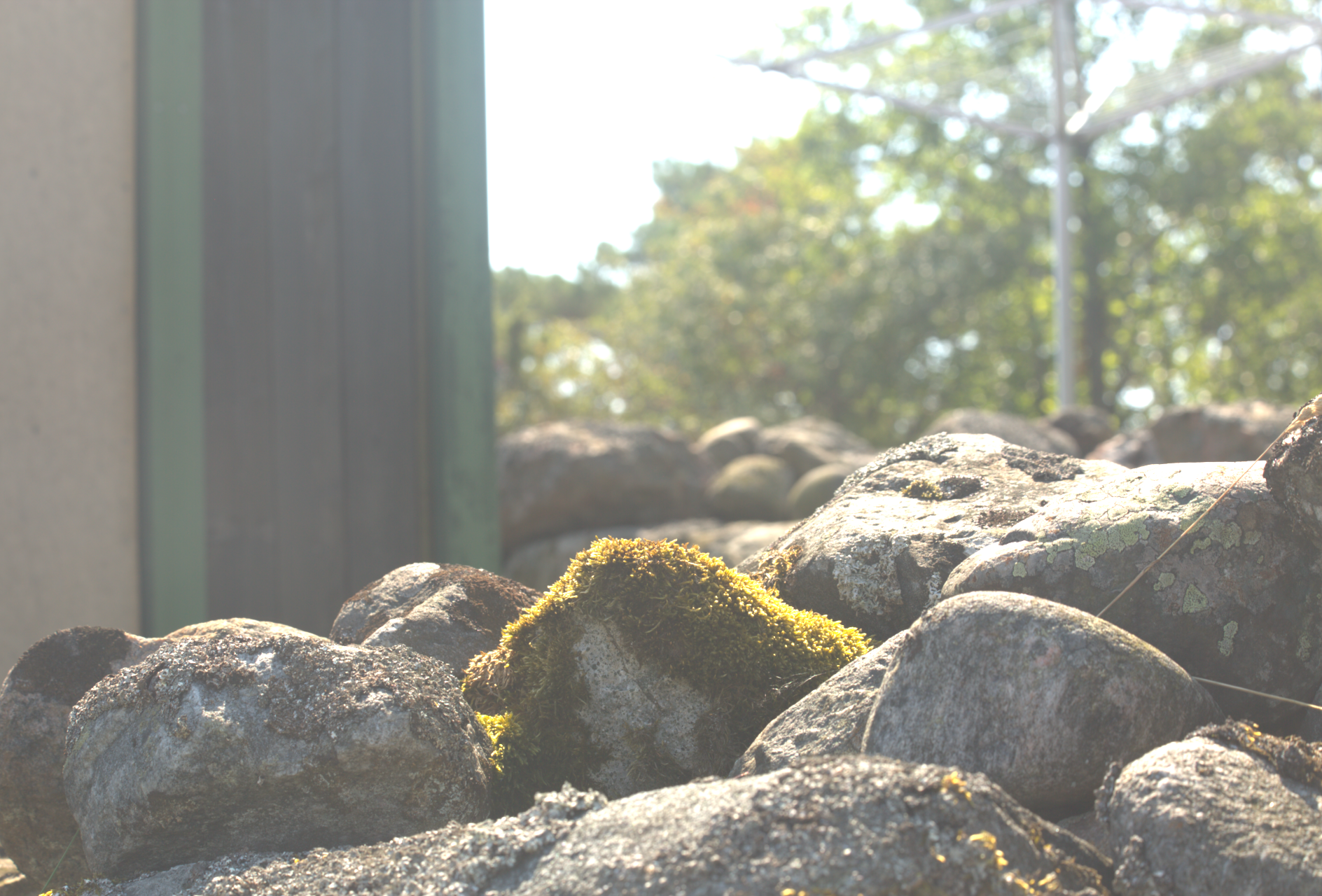}}\hskip.3em
\subfloat[]{\includegraphics[width=0.32\linewidth]{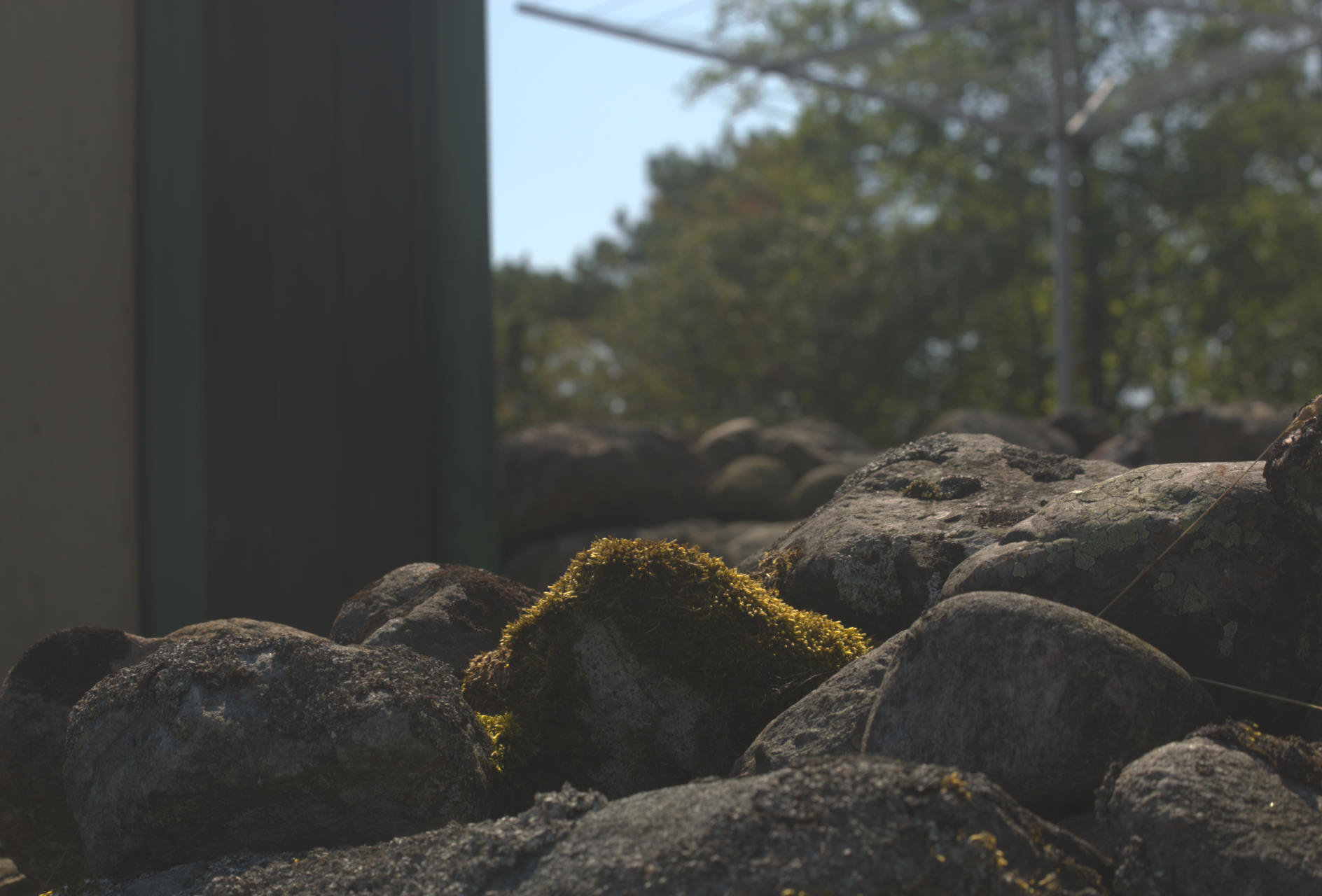}}\hskip.3em
\subfloat[]{\includegraphics[width=0.32\linewidth]{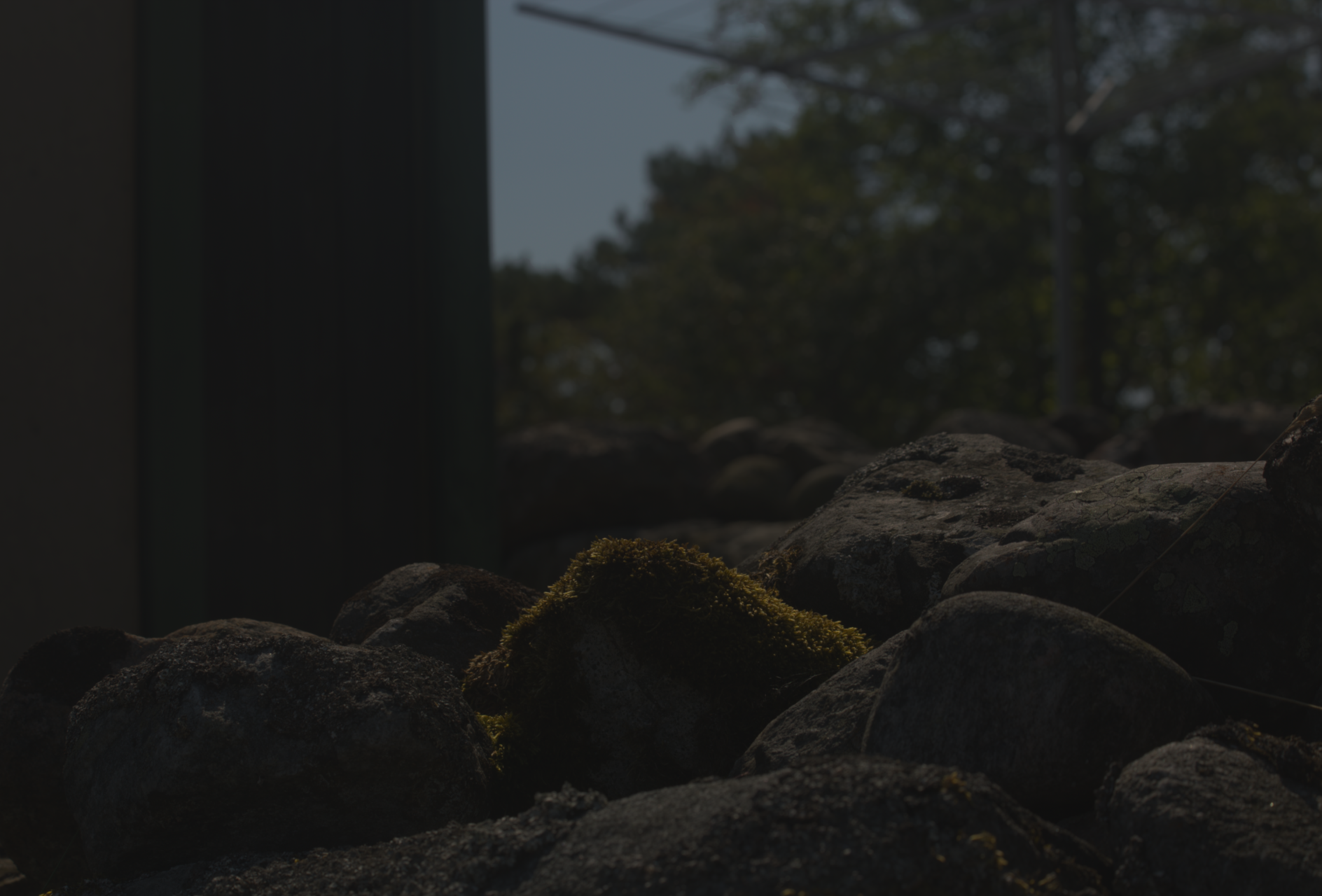}}
  \caption{Decomposition of an HDR image with eight stops into three LDR images of different exposures using the inverse display model in Eq.~\eqref{eq:IDM}. $l_0$ and $l_1$ denote the minimum and maximum luminances in the log-scale. \textbf{(a)} indicates the positions of the sliding windows by Eq.~\eqref{eq:position}.
  \textbf{(b)}-\textbf{(d)} are the LDR images corresponding to Window $1$ to Window $3$, respectively.
  }
  \label{fig:luminance}
  \vspace{-0.8em}
\end{figure}

We follow~\cite{mantiuk2009visualizing} to determine the positions of the sliding windows (\ie, the values of $\{v^{{(k)}}\}$). 
Specifically, we select $K$ uniformly spaced overlapping windows such that each eight stops\footnote{When photometric units (\eg, luminances) are plotted on the $\log_2$ axis, each logarithmic unit corresponds to $1$ stop.} of the luminance range are covered by three windows. This can be done by dividing the eight stops into three equal dynamic ranges and setting the \textit{endpoint} of the $k$-th window to be
\begin{align}\label{eq:position}
    l^{(k)} = {l}_{0}+\frac{8}{3}k,
\end{align}
where ${l}_{0}$ represents the minimum log-luminance in the scene. 
The exposure value $v^{(k)}$ is then computed by
\begin{align}\label{eq:ek}
    v^{(k)} = 2^{-l^{(k)}}.
\end{align}
Fig.~\ref{fig:luminance} (a) shows such an example HDR image with eight stops. 
Fig.~\ref{fig:luminance} (b)-(d) show the corresponding LDR images, which exhibit different exposures.

\begin{figure}[t]
  \centering
\subfloat{\includegraphics[width=0.32\linewidth]{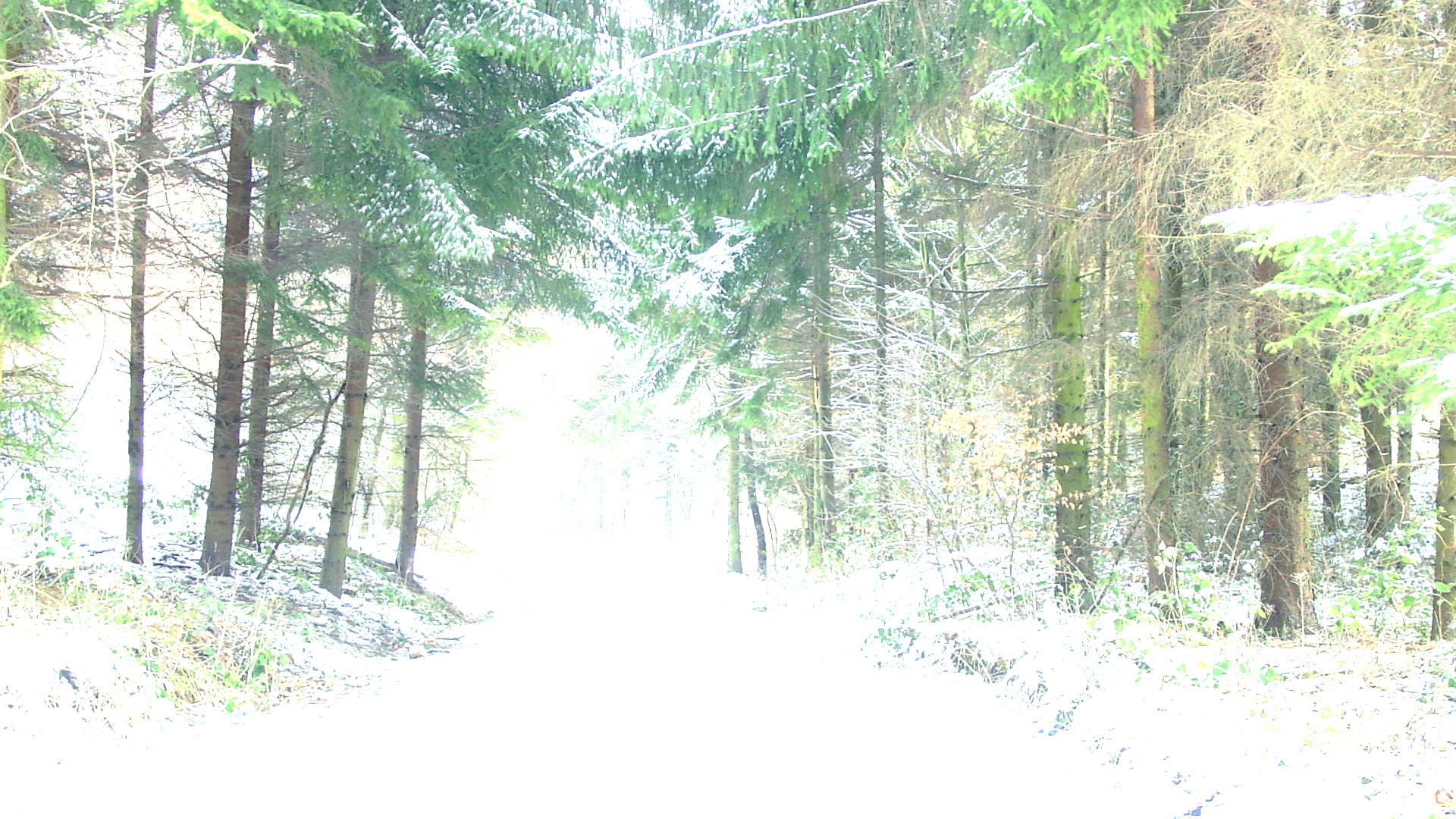}}\hskip.3em
\addtocounter{subfigure}{-1}
\subfloat[LDR images]{\includegraphics[width=0.32\linewidth]{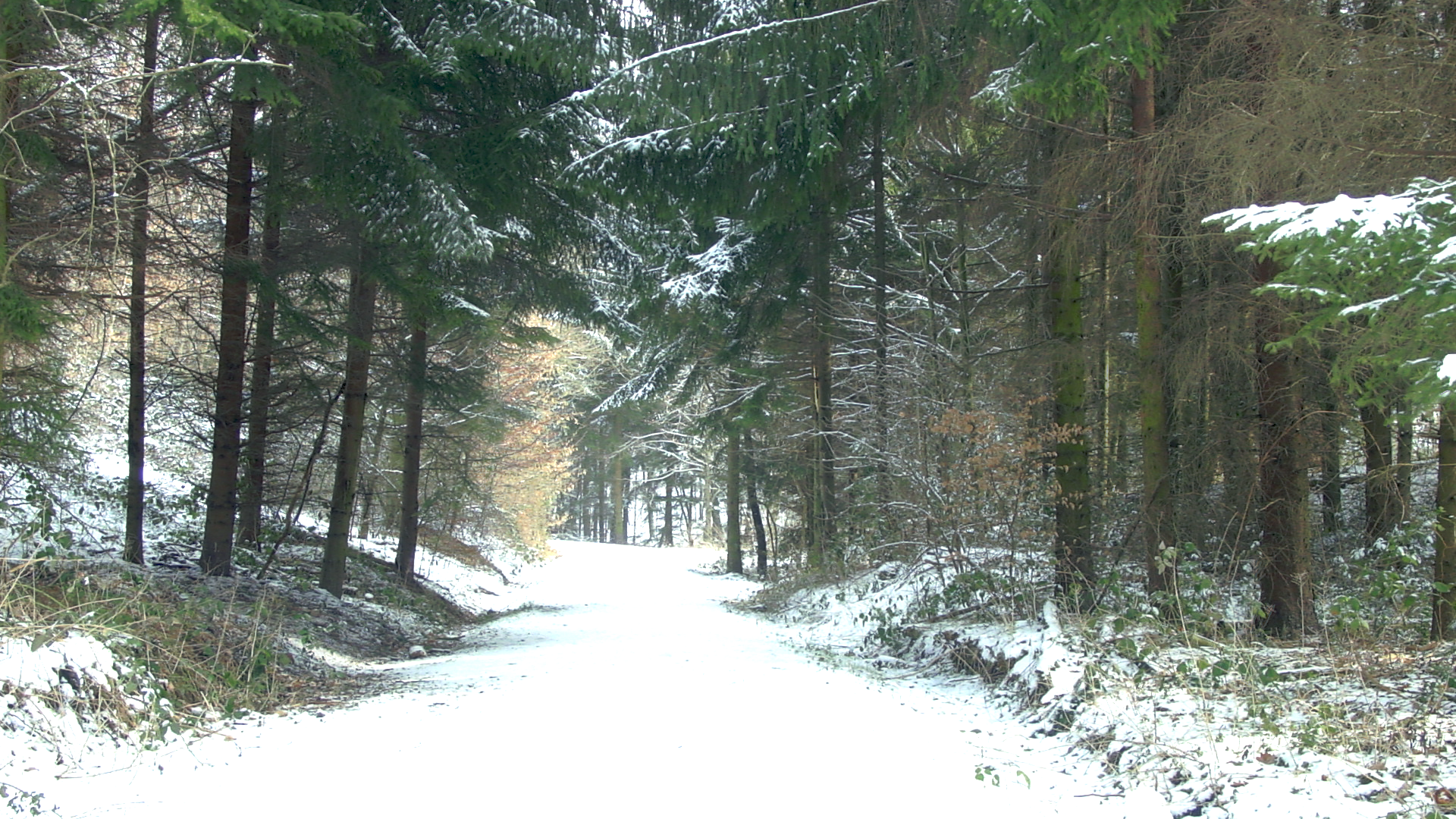}}\hskip.3em
\subfloat{\includegraphics[width=0.32\linewidth]{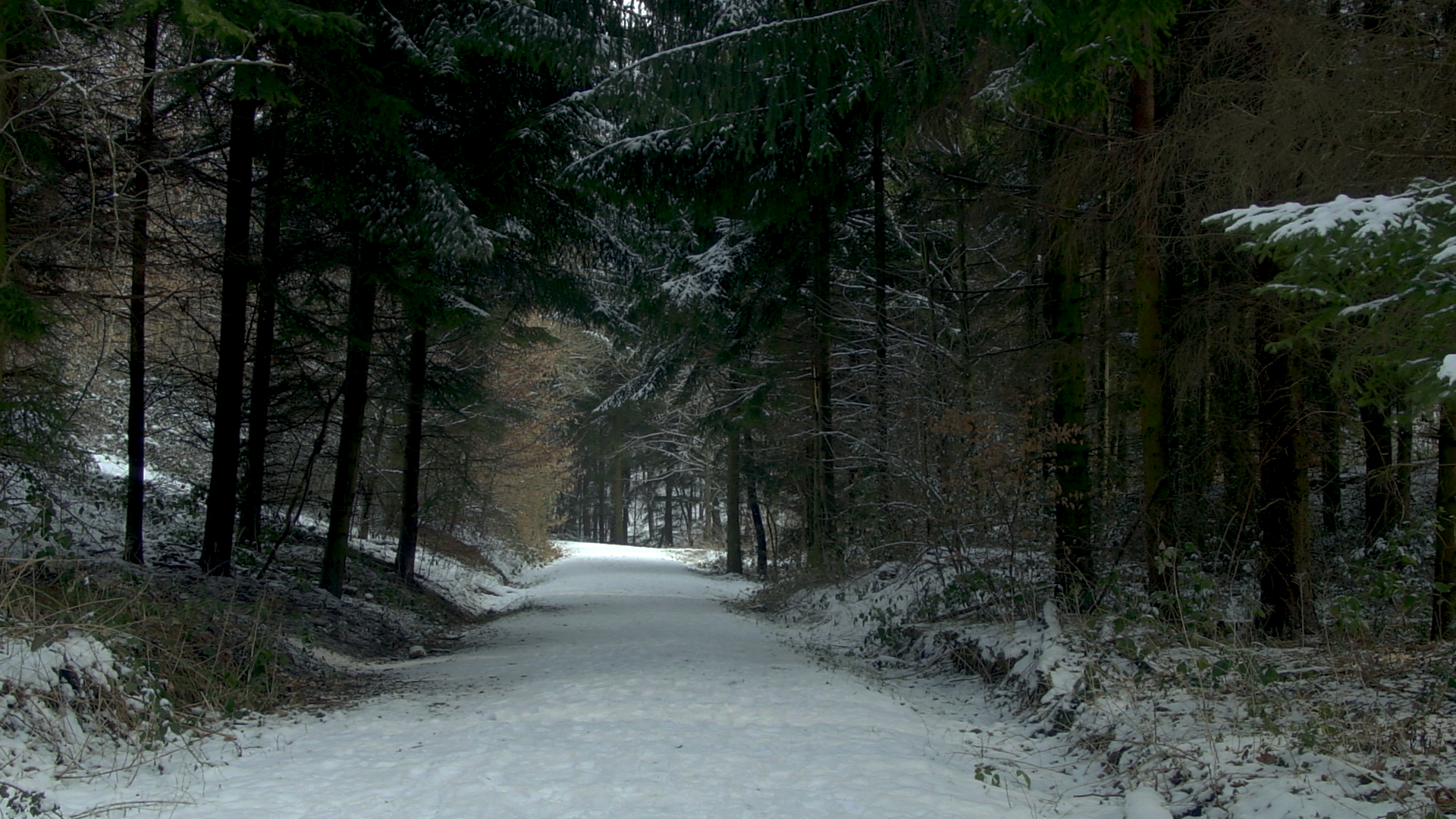}}
\\
\subfloat{\includegraphics[width=0.32\linewidth]{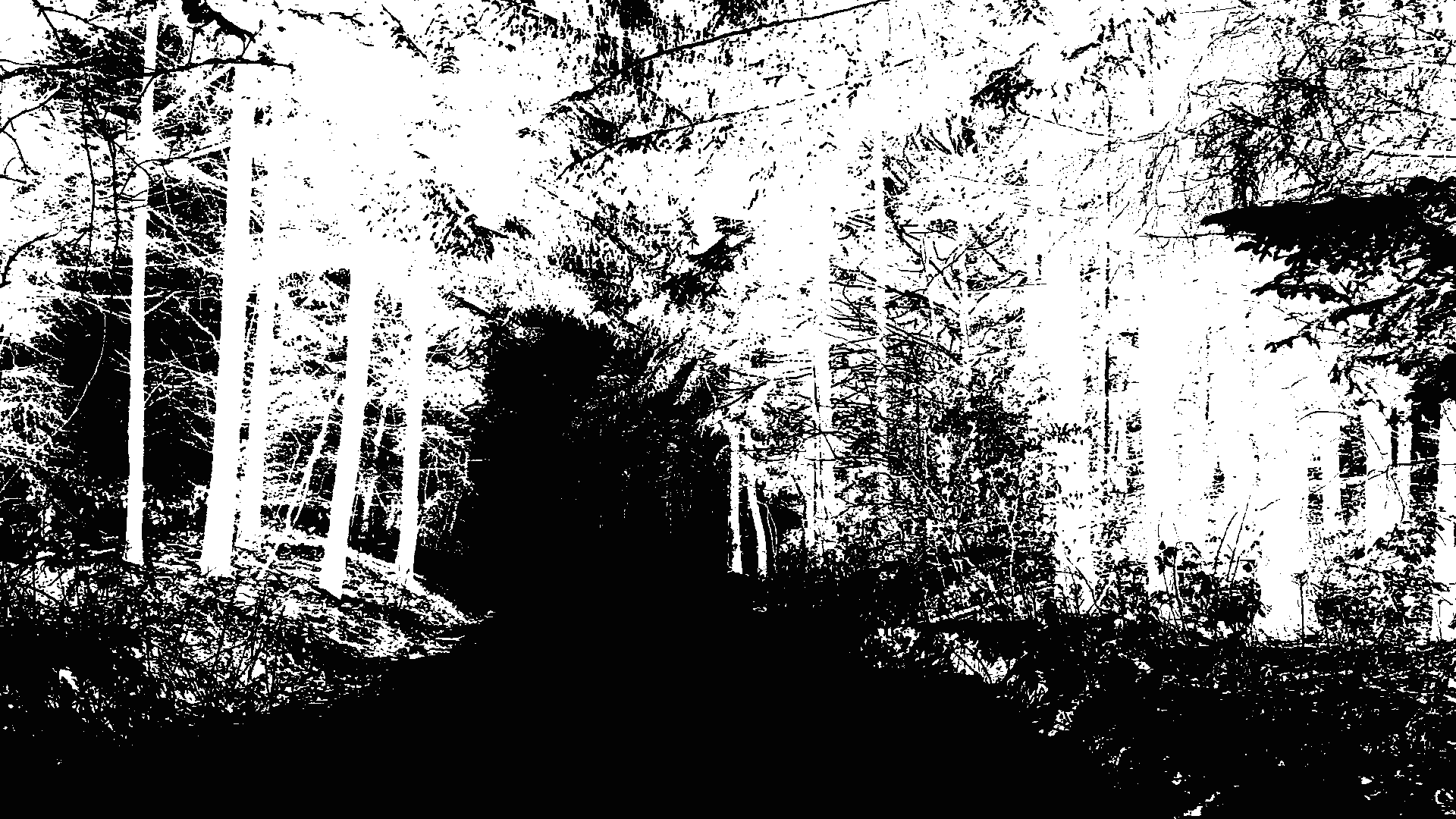}}\hskip.3em
\addtocounter{subfigure}{-2}
\subfloat[Weighting maps]{\includegraphics[width=0.32\linewidth]{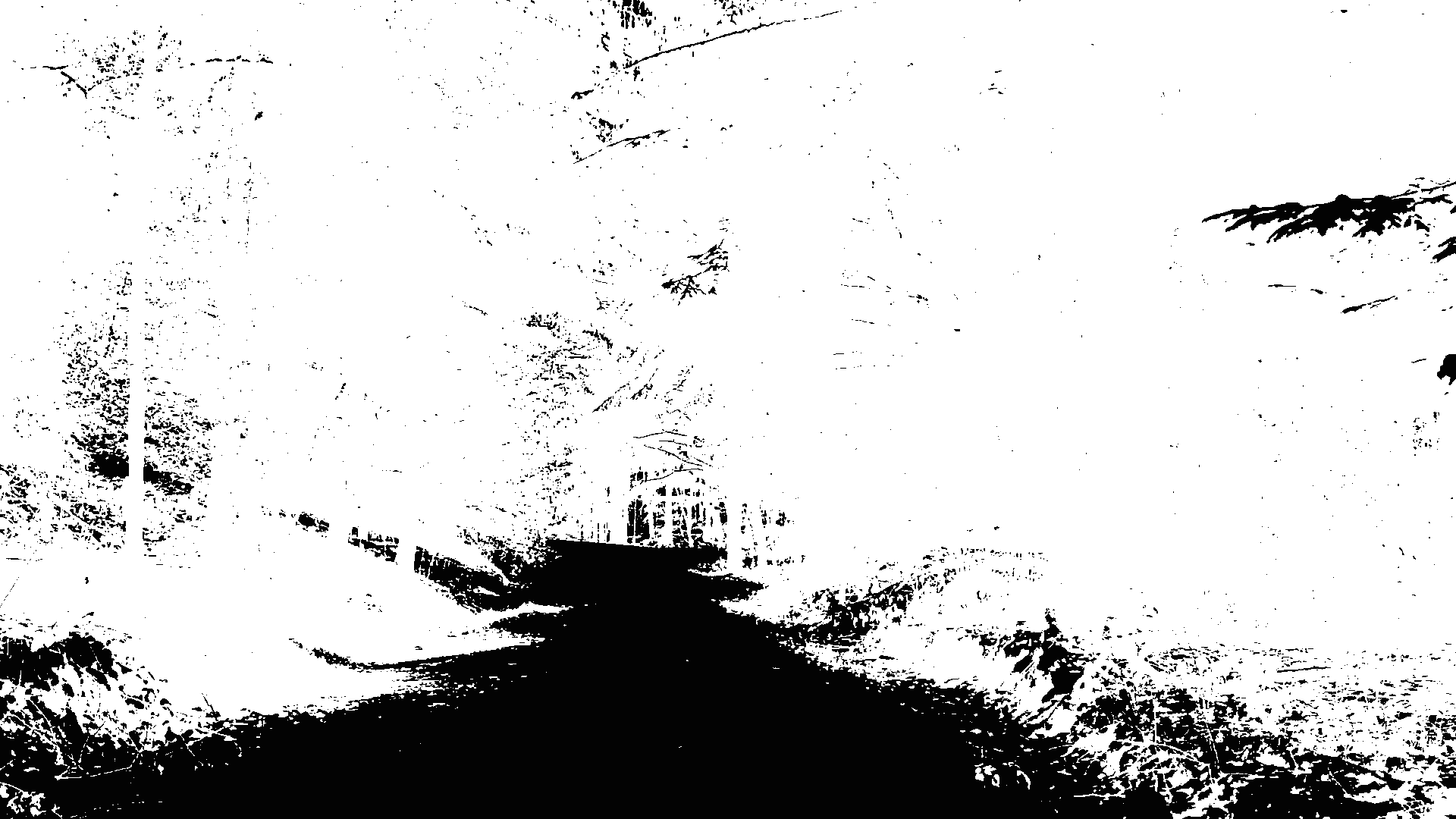}}\hskip.3em
\subfloat{\includegraphics[width=0.32\linewidth]{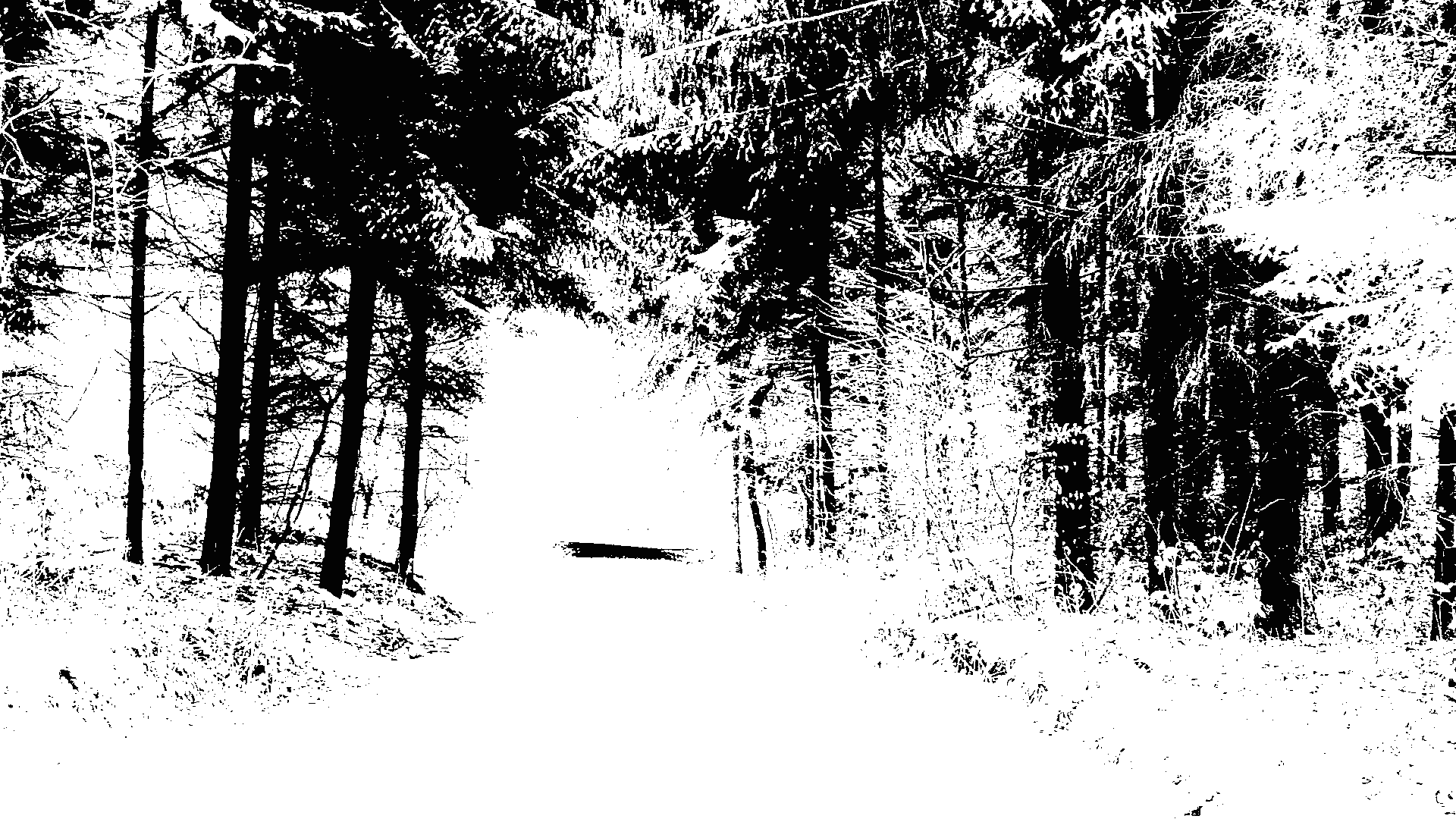}}
  \caption{LDR image stack generated by the inverse display model (in Eq.~\eqref{eq:IDM}) and the corresponding local weighting maps (by Eq.~\eqref{eqn:weight}) for the ``Forest" scene.}
  \label{fig:weight}
  \vspace{-0.8em}
\end{figure}

\subsection{Quality Assessment Model}
In the same vein, we may utilize another set of $\{\hat{v}^{(k)}\}^K_{k=1}$ to compute an LDR image stack $\{\hat{L}^{(k)}\}^K_{k=1}$ from the test HDR image $\hat{H}$.
Subsequently, we evaluate the perceptual quality of $\hat{L}^{(k)}$ using ${L}^{(k)}$ as reference:
\begin{align}\label{eq:basemodel}
    Q_i^{(k)}= D_i\left(L^{(k)},\hat{L}^{(k)};v^{(k)},\hat{v}^{(k)}\right),
\end{align}
where $D(\cdot,\cdot)$ denotes an LDR quality metric that produces a local quality map, indexed by $i$. A larger $Q_i^{(k)}$ indicates higher predicted quality at the $i$-th spatial location and $k$-th exposure. We pool local quality scores with a local weighting map: 
\begin{align}\label{eq:qk}
     Q^{(k)} = \frac{\sum_iW_i^{(k)} Q_i^{(k)}\left(L^{(k)},\hat{L}^{(k)};v^{(k)},\hat{v}^{(k)}\right)}{\sum_iW_i^{(k)}},
\end{align}
 where 
\begin{equation}
\label{eqn:weight}
W_i^{(k)} = \begin{cases}
1 & \mbox{if}\,\ 0.1\le L^{(k)}_i\le 0.9 \\
\epsilon & \mbox{otherwise},
\end{cases}
\end{equation}
is determined by a simple well-exposedness measure to exclude under- and over-exposed regions. $\epsilon$ is a small positive constant set to $10^{-5}$. In practice, we further normalize the local weightings for the same spatial location across different exposures (\ie, $\sum_k W^{(k)}_i = 1$) to make each spatial location contributes equally in the computation. Fig.~\ref{fig:weight} shows the local weighting maps corresponding to the LDR image stack of the ``Forest'' scene. The overall quality score is computed by aggregating global quality estimates across exposures:
\begin{align}\label{eq:gw}
     Q = \sum_{k=1}^{K}G^{(k)}Q^{(k)}\left(L^{(k)},\hat{L}^{(k)};v^{(k)},\hat{v}^{(k)}\right),
\end{align}
where $G^{(k)}$ is the $k$-th global weighting constrained to be non-negative, and $\sum_k G^{(k)}=1$. It is flexible to put more emphasis on assessing specific luminance ranges by raising the associated $G^{(k)}$ values. Unless otherwise specified, we set $G^{(k)} = 1/K$.

\setcounter{subfigure}{0}
\begin{figure}[!t]

\centering
\subfloat{\includegraphics[width=0.32\linewidth]{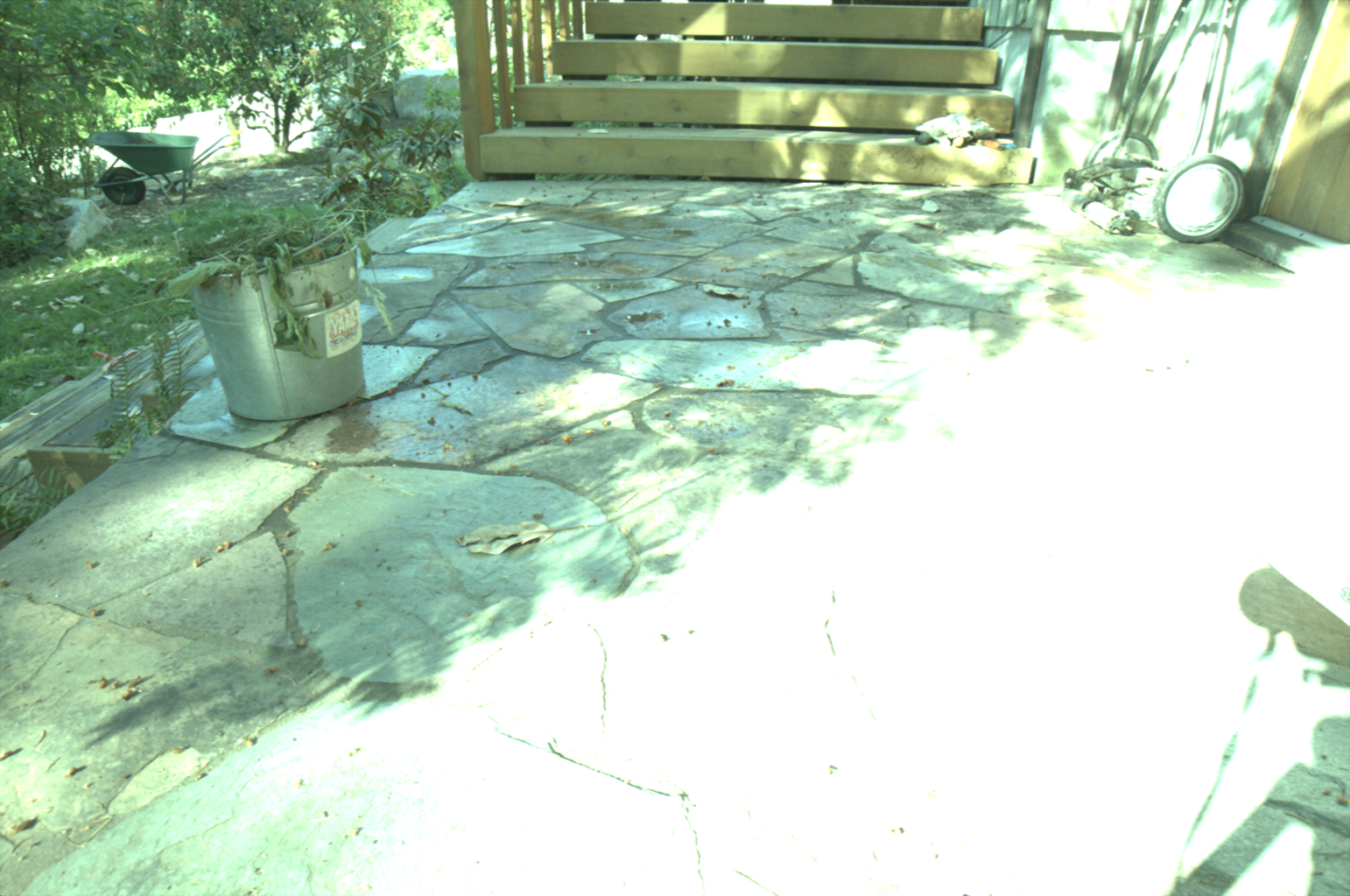}}\hskip.3em
\addtocounter{subfigure}{-1}
\subfloat[Reference]{\includegraphics[width=0.32\linewidth]{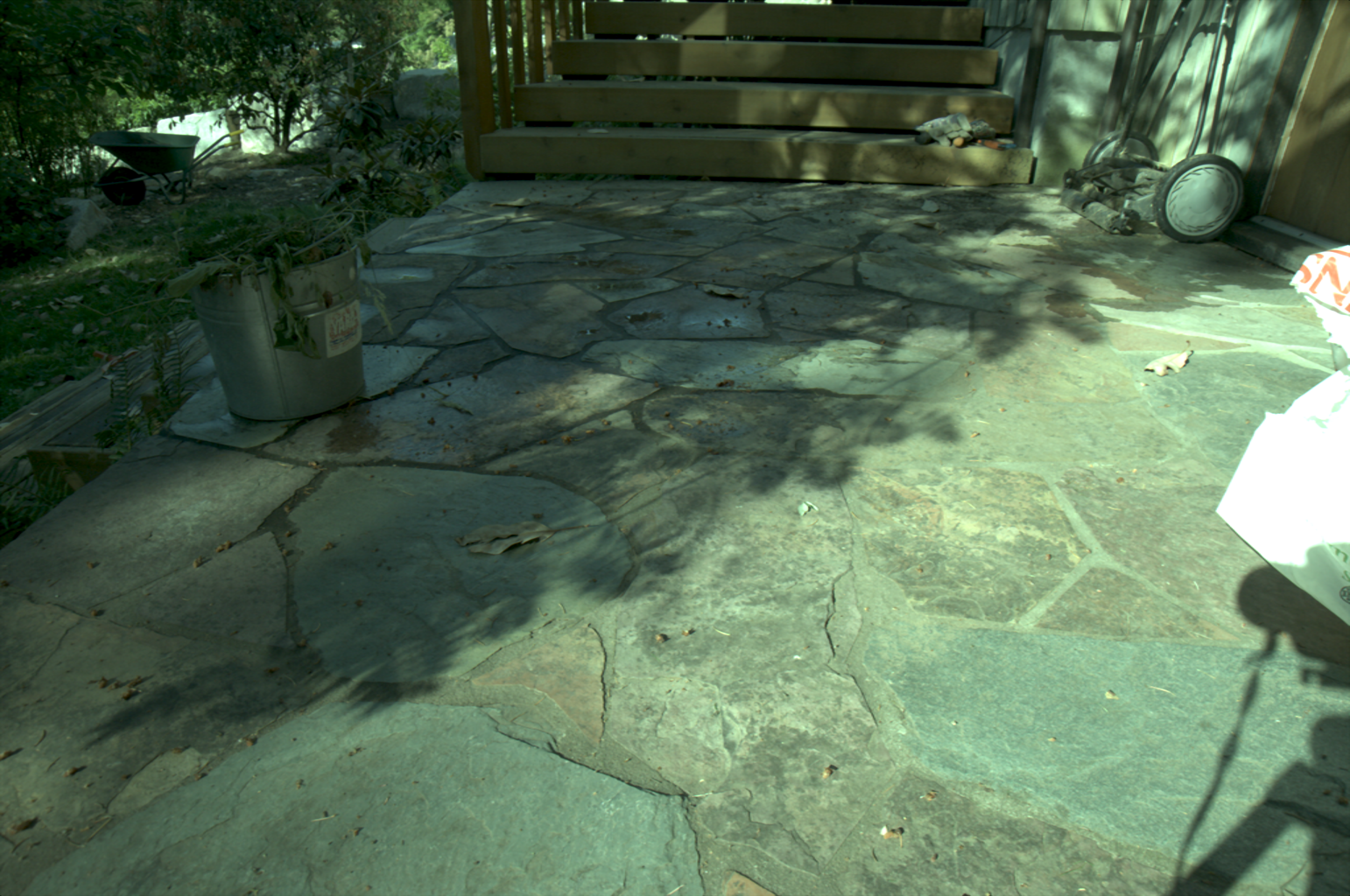}}\hskip.3em
\subfloat{\includegraphics[width=0.32\linewidth]{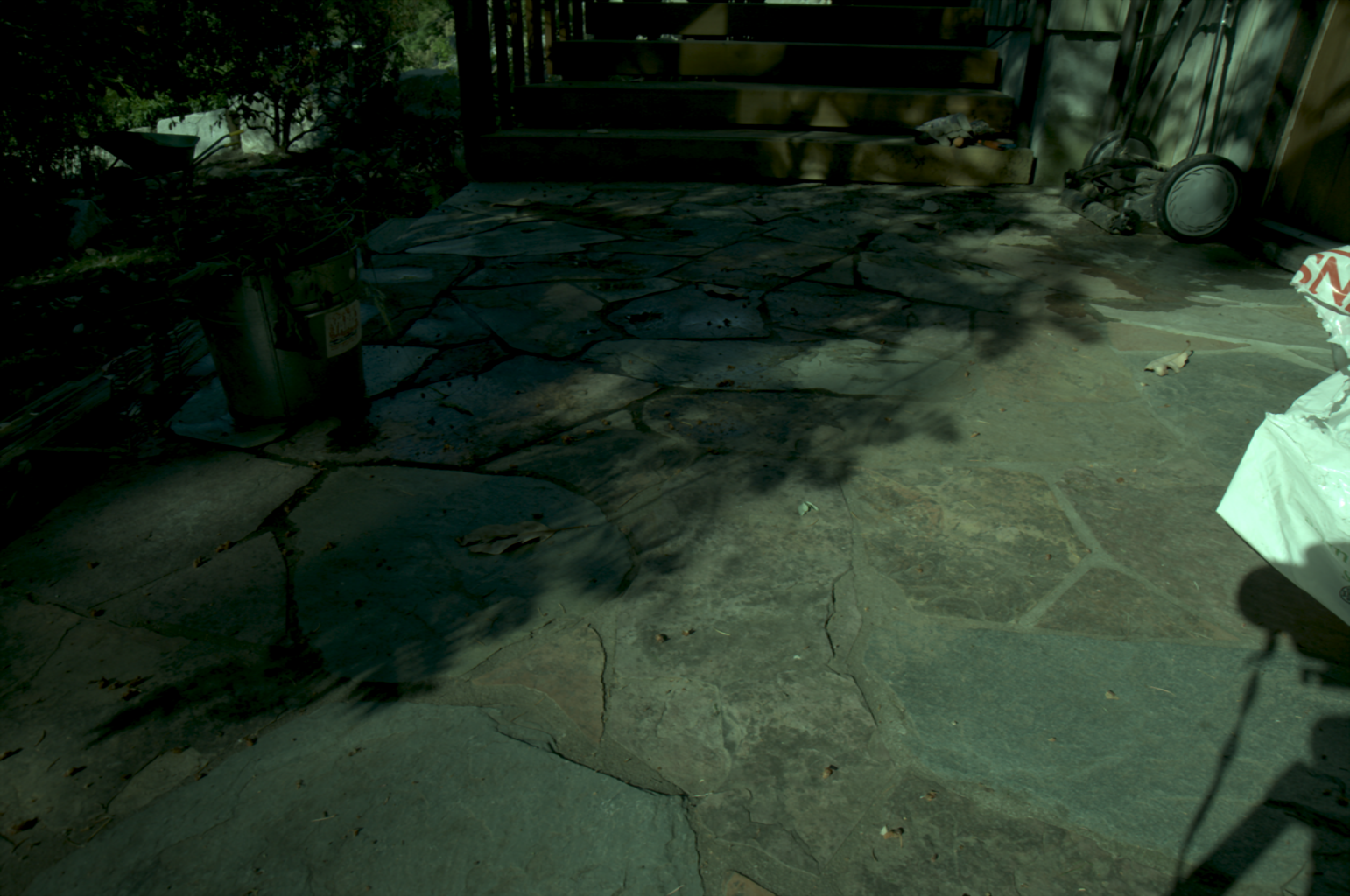}}
\\
\subfloat{\includegraphics[width=0.32\linewidth]{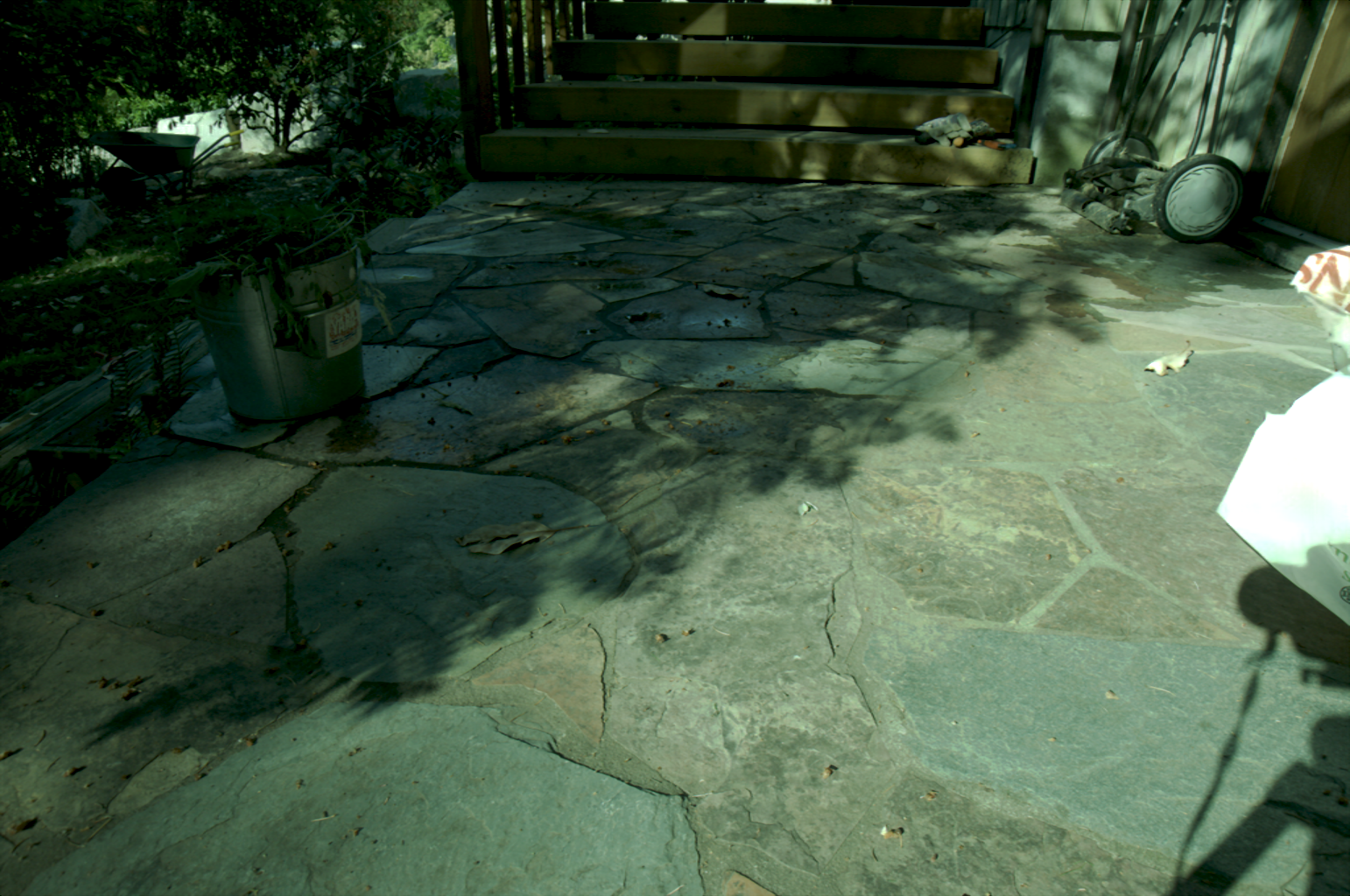}}\hskip.3em
\addtocounter{subfigure}{-2}
\subfloat[Uncompensated
]{\includegraphics[width=0.32\linewidth]{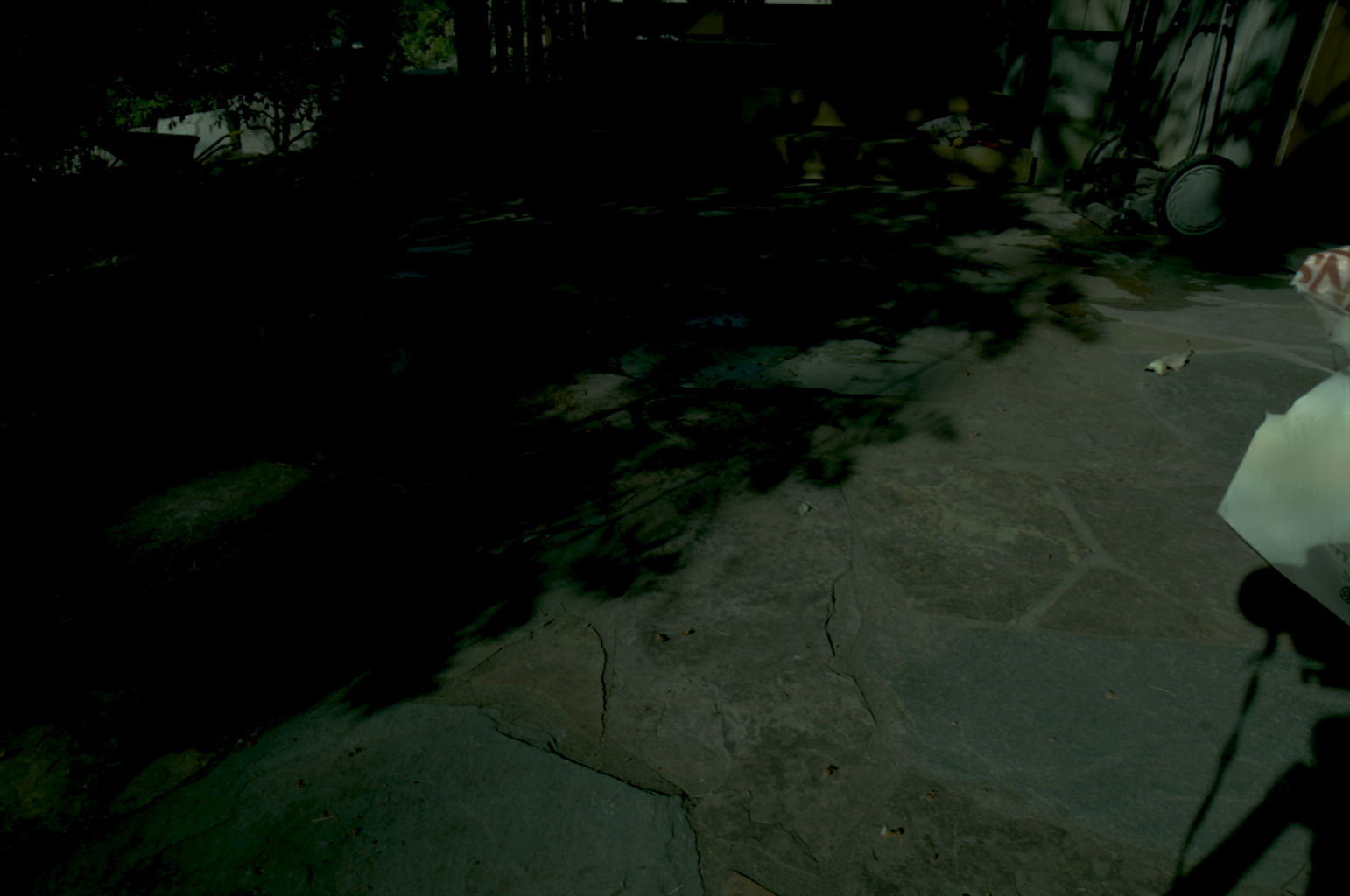}}\hskip.3em
\subfloat{\includegraphics[width=0.32\linewidth]{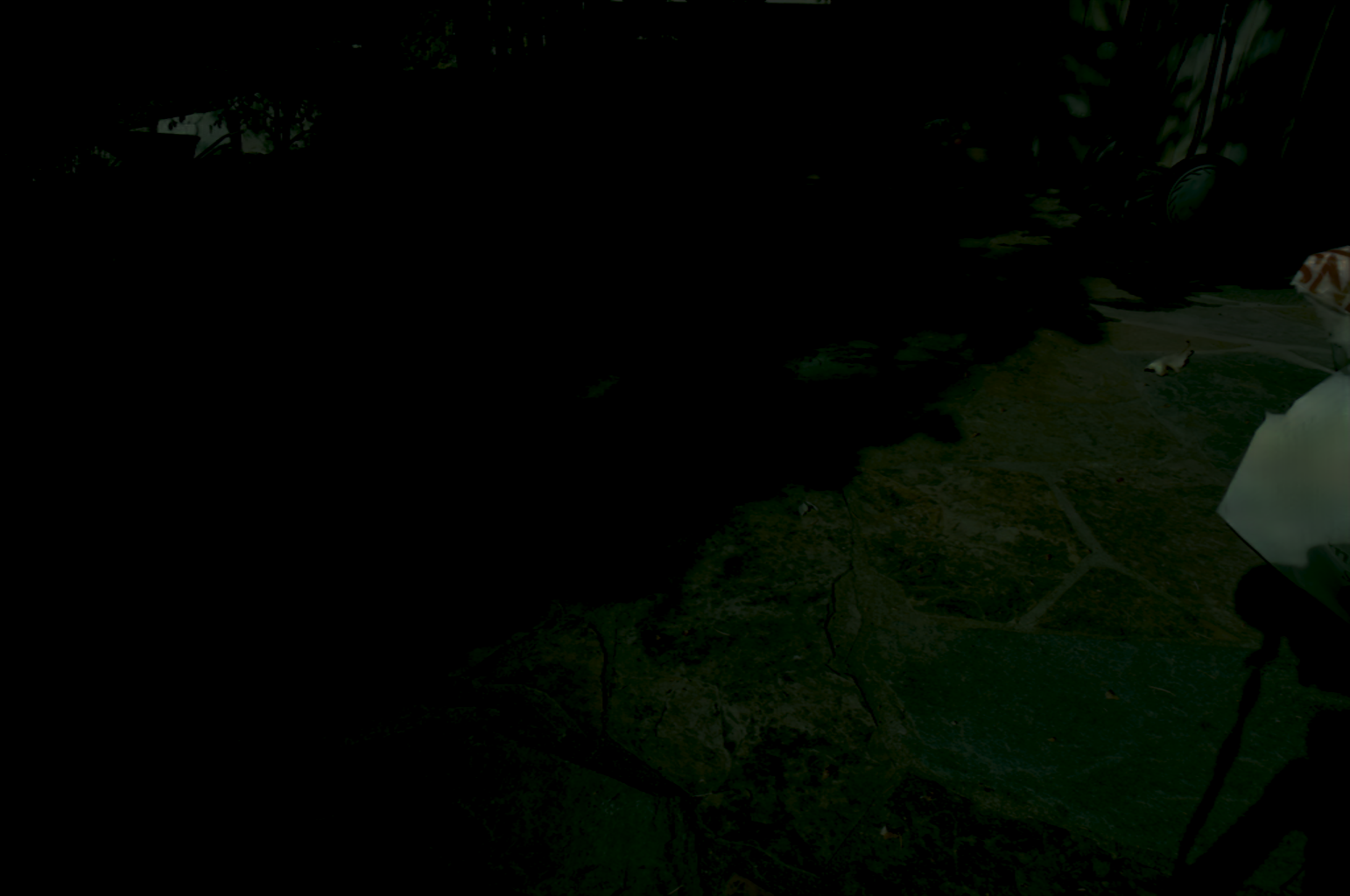}}
\\
\addtocounter{subfigure}{-2}
\subfloat{\includegraphics[width=0.32\linewidth]{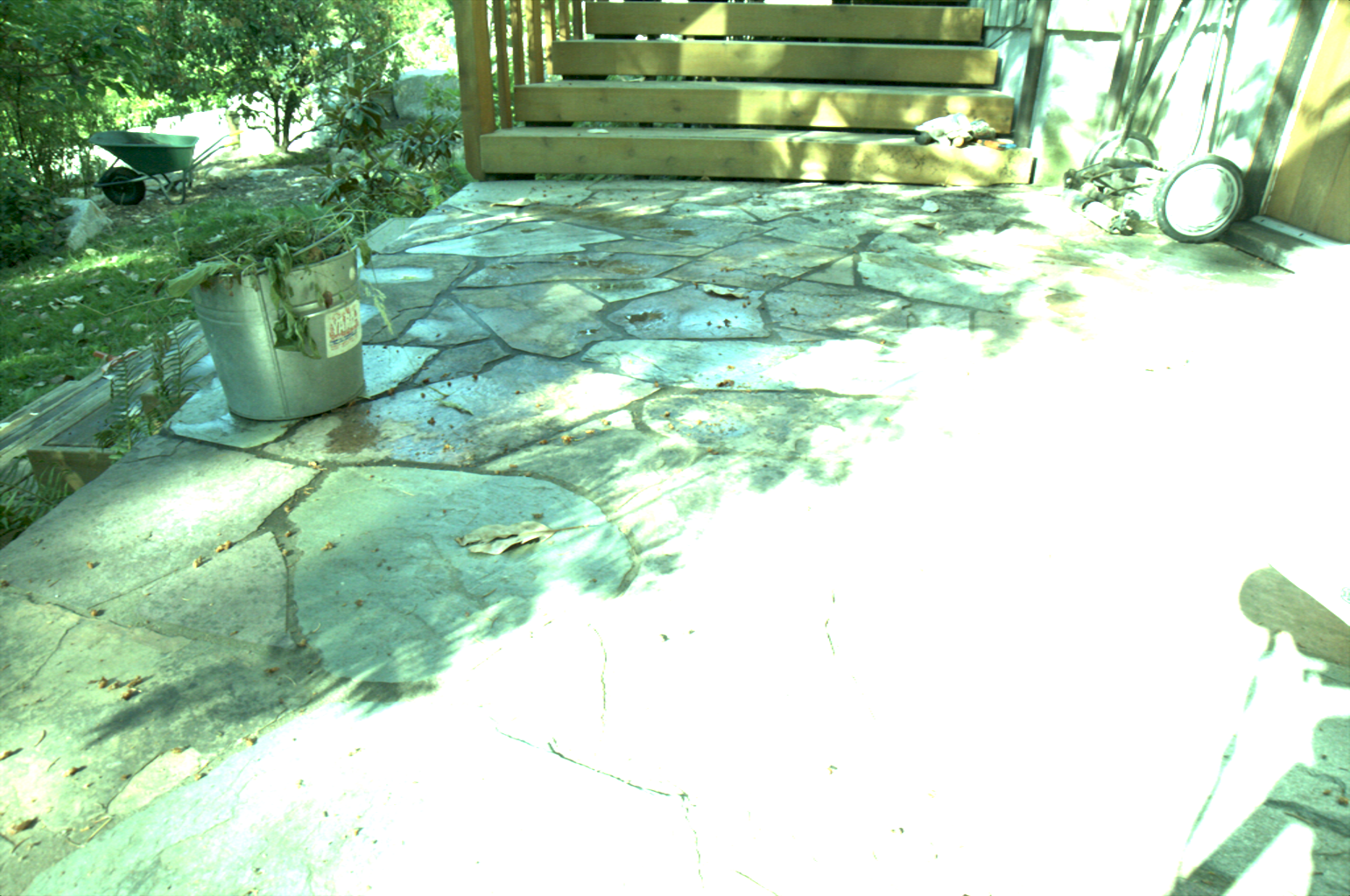}}\hskip.3em
\subfloat[Compensated]{\includegraphics[width=0.32\linewidth]{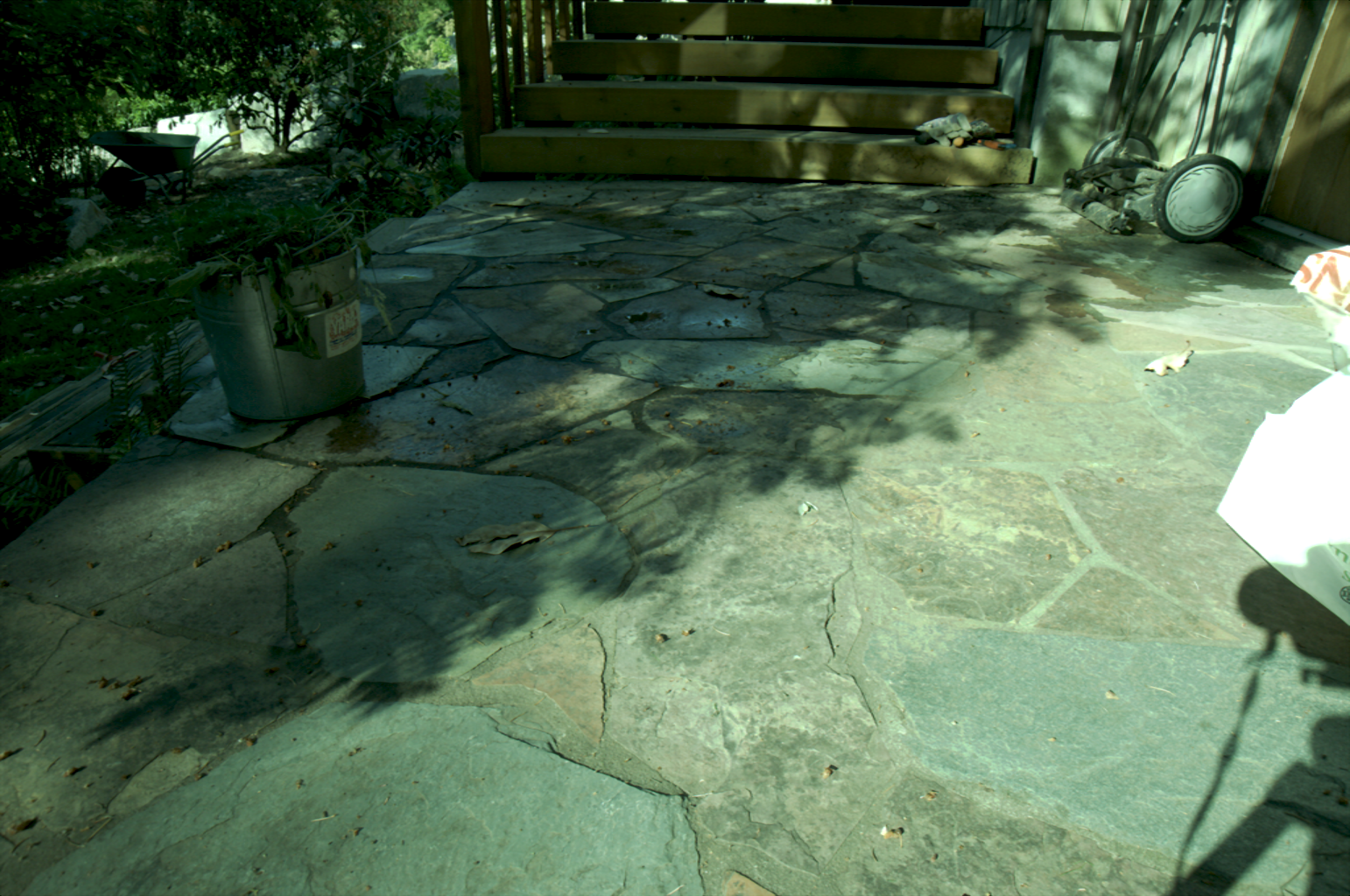}}\hskip.3em
\subfloat{\includegraphics[width=0.32\linewidth]{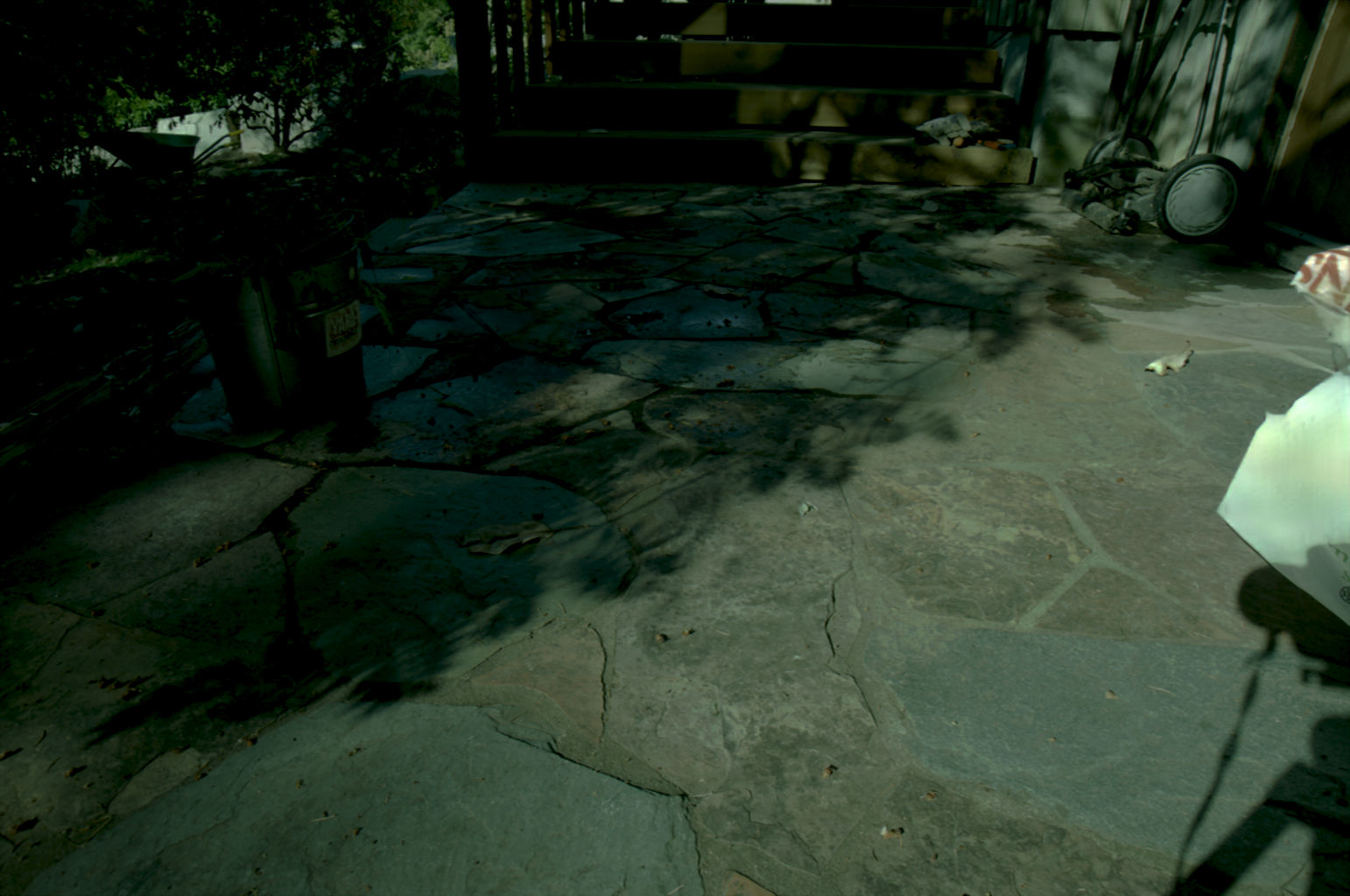}}
  \caption{Illustration of compensation for the luminance shifts through Eq.~\eqref{eq:omeh}. \textbf{(a)} LDR image stack generated from the reference HDR image. \textbf{(b)} LDR image stack generated from the HDR image by MaskHDR~\cite{Santos2020} \textit{with the same exposure values used in (a)}. \textbf{(c)} LDR image stack generated from the HDR image by MaskHDR~\cite{Santos2020} with optimized exposure values.}
  \label{fig:calibrated}
  \vspace{-0.8em}
\end{figure}

\begin{table*}[!t]
\centering
\caption{Performance comparison in terms of SRCC and PLCC of the proposed HDR quality metrics against $19$ existing methods on four HDR-IQA datasets. The left and right numbers separated by  ``/"  indicate the performance on the whole UPIQ dataset and its HDR image subset, respectively. The weightings to compute the average results in the last column are proportion to the numbers of HDR images in respective datasets. The top-$2$ results are highlighted in bold.}  \label{tab:hdr_metric}
\begin{threeparttable}
\resizebox{\linewidth}{!}{
\begin{tabular}{llcccccccccc}
\toprule
 \multirow{2}{*}{Model} & \multirow{2}{*}{}  & \multicolumn{2}{c}{Narwaria2013~\cite{narwaria2013tone}} & \multicolumn{2}{c}{Valenzise2014~\cite{valenzise2014performance}} & \multicolumn{2}{c}{Zerman2017~\cite{zerman2017extensive}} & \multicolumn{2}{c}{UPIQ~\cite{mikhailiuk2021consolidated}}& \multicolumn{2}{c}{Weighted Avg } \\
  &  & SRCC  & PLCC & SRCC & PLCC & SRCC & PLCC & SRCC & PLCC& SRCC & PLCC  \\
\hline
\multicolumn{2}{l}{NLPD}& 0.716 & 0.747 & 0.828 & 0.845 & 0.752 & 0.755 & 0.817/0.814 & 0.833/0.821& 0.785 & 0.797 \\
\multicolumn{2}{l}{HDR-VQM}& 0.761 & 0.788 & 0.865 & 0.892 & 0.762 & 0.774 & 0.788/0.818 & 0.817/0.819 & 0.801 & 0.812 \\
\multicolumn{2}{l}{HDR-VDP-3-$Q$}& 0.742 & 0.770 & 0.835 & 0.874 & 0.700 & 0.695 & 0.826/\textbf{0.845} & 0.871/\textbf{0.843} & 0.801 & 0.808 \\
\multicolumn{2}{l}{HDR-VDP-3-$D$}& 0.723 & 0.735 & 0.829 & 0.854 & 0.700 & 0.708 & 0.801/0.800 & 0.806/0.784 & 0.771 & 0.768 \\
\hline
\multicolumn{2}{l}{MAE}  & 0.107 & 0.101 & 0.225 & 0.373 & 0.361 & 0.355 & 0.534/0.256 & 0.570/0.294 & 0.238 & 0.269\\
\multicolumn{2}{l}{PU-MAE} & 0.543 & 0.620 & 0.435 & 0.625 & 0.553 & 0.568 & 0.556/0.620 & 0.602/0.639 & 0.579 & 0.623 \\ 
\multicolumn{2}{l}{PU21-MAE} & 0.560 & 0.600 & 0.470 & 0.575 & 0.557 & 0.555 & 0.585/0.613 & 0.625/0.617 & 0.582 & 0.601 \\
\hline
\multicolumn{2}{l}{$Q^\star_\textrm{MAE}$} & 0.624 & 0.653 & 0.838 & 0.868 & 0.750 & 0.728 & 0.602/0.642 & 0.635/0.646 & 0.670 & 0.677 \\        
\hline
\multicolumn{2}{l}{PSNR}  & 0.124 & 0.139 & 0.371 & 0.416 & 0.465 & 0.506 & 0.645/0.299 & 0.650/0.341 & 0.293 & 0.329 \\
\multicolumn{2}{l}{PU-PSNR} & 0.532 & 0.595 & 0.529 & 0.611 & 0.649 & 0.677 & 0.643/0.631 & 0.651/0.644 & 0.605 & 0.636\\ 
\multicolumn{2}{l}{PU21-PSNR} & 0.546 & 0.574 & 0.588 & 0.655 & 0.633 & 0.662 & 0.666/0.585 & 0.665/0.591 & 0.584 & 0.603 \\
\hline
\multicolumn{2}{l}{$Q^\star_\textrm{PSNR}$} & 0.682 & 0.716 & 0.766 & 0.812 & 0.789 & 0.774 & 0.700/0.709 & 0.701/0.716 & 0.720 & 0.733\\        
\hline
\multicolumn{2}{l}{SSIM} & 0.126 & 0.313 & 0.322 & 0.502  & 0.493 & 0.451 & 0.677/0.383 & 0.706/0.475 & 0.341 & 0.440\\
\multicolumn{2}{l}{PU-SSIM} & 0.651 & 0.690 & 0.840 & 0.880 & 0.754 & 0.750 & 0.665/0.736 & 0.667/0.738 & 0.729 & 0.741\\
\multicolumn{2}{l}{PU21-SSIM} & 0.633 & 0.679 & 0.837 & 0.863 & 0.757 & 0.744 & 0.680/0.674 & 0.677/0.671 & 0.691 & 0.699 \\
\hline
\multicolumn{2}{l}{$Q^\star_\textrm{SSIM}$} & 0.658 & 0.664 & 0.893 & 0.917 & 0.814 & 0.801 & 0.731/0.750 & 0.745/0.747 & 0.752 & 0.751\\ 
\hline
\multicolumn{2}{l}{LPIPS} & 0.650 & 0.695 & 0.768 & 0.780 & 0.684 & 0.695 & 0.844/0.829 & 0.876/0.824 & 0.765 & 0.774 \\
\multicolumn{2}{l}{PU-LPIPS} & 0.801 & 0.823 & 0.883 & 0.922 & 0.779 & 0.759 & 0.834/0.832 & 0.870/0.837 & 0.822 & 0.829\\
\multicolumn{2}{l}{PU21-LPIPS} & 0.815 & 0.833 & 0.903 & 0.921 & 0.806 & 0.804 & 0.779/0.822 & 0.838/0.828 & 0.825 & 0.833\\
\hline
\multicolumn{2}{l}{$Q^\star_\textrm{LPIPS}$} & 0.823 & 0.839 & 0.905 & 0.918 & 0.847 & 0.837 & 0.844/0.836 & 0.880/0.835 & \textbf{0.840} & \textbf{0.843} \\
\hline

\multicolumn{2}{l}{DISTS} & 0.515 & 0.593 & 0.794 & 0.848 & 0.811 & 0.846 & \textbf{0.860}/0.691 & \textbf{0.882}/0.701 & 0.680 & 0.712\\
\multicolumn{2}{l}{PU-DISTS} & 0.847 & 0.867 & \textbf{0.910} & \textbf{0.929} &\textbf{0.862} & \textbf{0.870} &  0.805/0.788 & 0.857/0.804 & 0.821 & 0.837\\
\multicolumn{2}{l}{PU21-DISTS} & \textbf{0.860} & \textbf{0.872} & 0.907 & 0.921 & 0.829 & 0.831 & 0.798/0.801 & 0.854/0.822 & 0.826 & 0.842\\
\hline
\multicolumn{2}{l}{$Q^\star_\textrm{DISTS}$} & \textbf{0.868} & \textbf{0.877} & \textbf{0.917} & \textbf{0.930} & \textbf{0.904} & \textbf{0.901} & \textbf{0.861}/\textbf{0.853} & \textbf{0.881}/\textbf{0.857}& \textbf{0.869} & \textbf{0.873} \\

\bottomrule
\end{tabular}}
    \end{threeparttable}
    \vspace{-0.8em}
\end{table*}

As noticed by Hanji~\etal~\cite{Hanji2022}, the reference and test HDR images may exhibit luminance shifts that will significantly bias quality prediction. To mitigate this issue, we opt to further maximize $Q$
in Eq.~\eqref{eq:gw} with respect to $\{\hat{v}^{(k)}\}_{k=1}^K$: 
\begin{align}\label{eq:omeh}
     Q^{\star}= \max_{\{\hat{v}^{(k)}\}_{k=1}^{K}}Q\left(\{L^{(k)},\hat{L}^{(k)};v^{(k)},\hat{v}^{(k)}\}_{k=1}^K\right),
\end{align}
which can be decomposed into $K$ one-dimensional optimization problems, and solved efficiently by the golden-section, bisection, or Newton's methods. Fig.~\ref{fig:calibrated} provides a visual comparison of the test LDR image stacks without and with the luminance shift compensation. When adopting $Q$ in Eq.~\eqref{eq:gw} as the loss function for perceptual optimization of HDR image rendering tasks, we can more effectively minimize the luminance shifts in an online fashion by setting  $\hat{v}^{(k)} = v^{(k)}$, for $1\le k\le K$.

The proposed HDR quality metric naturally reduces to its base LDR metric ($D(\cdot)$ in Eq.~\eqref{eq:basemodel}) when assessing LDR images. This is because the LDR images would need to be first transformed from the display-encoded color space (\eg, sRGB) to linear color values through the forward display model:
\begin{align}\label{eq:mapping}
    L =  (1-b)P^{\gamma}+b\quad \mbox{and}\quad \hat{L} =  (1-b)\hat{P}^{\gamma}+b,
\end{align}
where $P$ and $\hat{P}$ are digital pixel values of the reference and test LDR images, respectively. The black-level factor $b$ and gamma parameter $\gamma$  are the same as in Eq.~\eqref{eq:IDM}. The maximum luminances of $L$ and $\hat{L}$ are scaled to $200$ $\rm cd/m^{2}$. The integration of the forward display model with the inverse display model in Eq.~\eqref{eq:IDM} results in an identity mapping, thereby leaving the input LDR image intact.

\section{Quality Assessment Validation} \label{sec:metric_expriment}
In this section, we compare our HDR quality metrics with existing model-based and encoding-based methods on four HDR-IQA datasets.

\subsection{Experimental Setups}
\noindent\textbf{Implementation Details}.
We adopt five base LDR quality models to implement $D(\cdot,\cdot)$ in Eq.~\eqref{eq:basemodel}: the mean absolute error (MAE), PSNR, SSIM, the learned perceptual image patch similarity (LPIPS) model~\cite{zhang2018unreasonable} with VGGNet~\cite{Simonyan14c}, and the deep image structure and texture similarity (DISTS) metric~\cite{ding2020image}. To solve the $K$ one-dimensional optimization problems in Eq.~\eqref{eq:omeh}, we employ the gradient ascent method with an initial learning rate of $10^{-3}$, and decay the learning rate by a factor of $5$ for every $1,000$ iterations with a maximum of $5,000$ iterations. Early stopping is enabled if the absolute difference of the losses between two consecutive iterations is less than $10^{-3}$. 

\noindent\textbf{Datasets.} Four publicly available HDR-IQA datasets are adopted for benchmarking: Narwaria2013~\cite{narwaria2013tone}, Valenzise2014~\cite{valenzise2014performance}, Zerman2017~\cite{zerman2017extensive}, and UPIQ~\cite{mikhailiuk2021consolidated}, which contain $140$, $50$, $100$ and $4,159$ images, respectively. The UPIQ dataset stands out for its collection of $380$ HDR and $3,779$ LDR images from four sub-datasets~\cite{narwaria2013tone, korshunov2015subjective, Sheikh2006Statistical, ponomarenko2015image}, whose scores have been carefully re-aligned to a common perceptual scale to ensure consistency.

\noindent\textbf{Competing Metrics}. We select nine model-based methods for comparison, including 1) NLPD~\cite{laparra2017perceptually}, 2) HDR-VQM~\cite{HDR-VQM}, 3) the quality score of HDR-VDP-3~\cite{mantiuk2023hdr} (denoted by HDR-VDP-3-$Q$) and 4) the difference score of HDR-VDP-3\footnote{The current version under evaluation is HDR-VDP-3.0.7 with the default parameter setting.} (denoted by HDR-VDP-3-$D$) as four HDR quality metrics, and 5) MAE, 6) PSNR, 7) SSIM~\cite{wang2004image}, 8) LPIPS~\cite{zhang2018unreasonable} and 9) DISTS~\cite{ding2020image} as five LDR quality metrics. We also equip the five LDR quality models with the PU~\cite{aydin2008extending} and PU21 encoding, giving rise to 10) PU-MAE, 11) PU-PSNR, 12) PU-SSIM, 13) PU-LPIPS, 14) PU-DISTS, 15) PU21-MAE, 16) PU21-PSNR, 17) PU21-SSIM, 18) PU21-LPIPS,  and 19) PU21-DISTS. 
As suggested in~\cite{aydin2008extending,Mantiuk2011hdrvdp2,azimi2021pu21}, we assume a test HDR display model with a maximum luminance of 
$1,000$ and $4,000$ $\rm cd/m^{2}$ for the PU and PU21 encoding, respectively, which are independently applied to the three color channels. We find empirically that the performance ranking is fairly robust to the selection of the maximum luminance of the display. For the LDR images in UPIQ, we first convert digital pixel values to luminance values via the display model in Eq.~\eqref{eq:mapping} before applying HDR quality metrics, and adjust the hyperparameters of the base LDR quality metrics if necessary\footnote{For example, in PU21-SSIM, the two normalizing constants are adjusted to $C_1 = (0.01\times 4000)^2 =1,600$ and $C_2 = (0.03\times 4000)^2 = 14,440$, respectively, as the maximum luminance is $4,000$ $\rm cd/m^{2}$.}. 

\noindent\textbf{Evaluation Criteria}. We use two evaluation criteria: Spearman's rank correlation coefficient (SRCC) and Pearson linear correlation coefficient (PLCC). As a standard practice~\cite{video2000final, Sheikh2006Statistical}, we fit a four-parameter logistic function before computing PLCC.

\begin{table}[!t]
  \centering
    \caption{Performance comparison in terms of SRCC and PLCC of the proposed HDR quality metrics without and with the luminance shift compensation.} \label{tab:optimization}
    
    \begin{tabular}{lccccc}
        \toprule 
        \multirow{3}{*}[-3pt]{Method} & \multicolumn{5}{c}{Weighted average across the four datasets}\\
        &\multicolumn{2}{c}{w/o compensation} & &\multicolumn{2}{c}{w/ compensation} \\ 
        \cmidrule{2-3}  \cmidrule{5-6} 
        & \textrm{SRCC} & \textrm{PLCC}
        & & \textrm{SRCC} & \textrm{PLCC} \\ \hline

        $Q_\textrm{MAE}$ & 0.525 & 0.586 & & 0.670 & 0.677 \\
        $Q_\textrm{PSNR}$ & 0.537 & 0.546 & & 0.720 & 0.733 \\
        $Q_\textrm{SSIM}$ & 0.575 & 0.600 & & 0.752 & 0.751 \\
        $Q_\textrm{LPIPS}$ & 0.694 & 0.702 & & 0.840 &0.843 \\
        $Q_\textrm{DISTS}$ & 0.708 & 0.711 & & 0.869 & 0.873 \\
        \bottomrule
    \end{tabular}
    \vspace{-0.8em}
\end{table}

\subsection{Results}
Table~\ref{tab:hdr_metric} presents the performance comparison results, where we find that the adopted inverse display model leads to consistent improvements for all base LDR quality models. In particular, the instantiation $Q^\star_\textrm{DISTS}$ achieves the best results on all four datasets, even surpassing the re-calibrated HDR-VDP-3-$Q$ on UPIQ. Consistent with previous studies~\cite{aydin2008extending,azimi2021pu21}, the PU and PU21 encoding can boost the performance of base LDR quality measures, but not as substantial as our metrics. When applied to the LDR images in UPIQ, the PU and PU21 encoding incur noticeable performance degradation. In stark contrast, our metrics maintain reliable LDR-IQA capabilities. Last, there is a clear trend that a better base LDR quality metric generally delivers better performance, affirming our objective of transferring the advancements in LDR-IQA to HDR-IQA.

Table~\ref{tab:optimization} shows the ablation results of the proposed HDR quality metrics without and with the luminance shift compensation (see Eq.~\eqref{eq:omeh}). It is evident that our compensated metrics consistently outperform the non-compensated counterparts. This performance gap is expected to be even more pronounced in the presence of large luminance shifts, such as assessing HDR images derived from single image HDR reconstruction methods~\cite{eilertsen2021cheat,Hanji2022}. Thus, compensating for luminance shifts is recommended as a standard procedure when comparing HDR images.

\section{Perceptual Optimization Validation}
In this section, we explore the application of the proposed HDR quality metrics for perceptual optimization of HDR novel view synthesis.

\begin{figure}[!t]
\centering
  \includegraphics[width=0.34\textwidth]{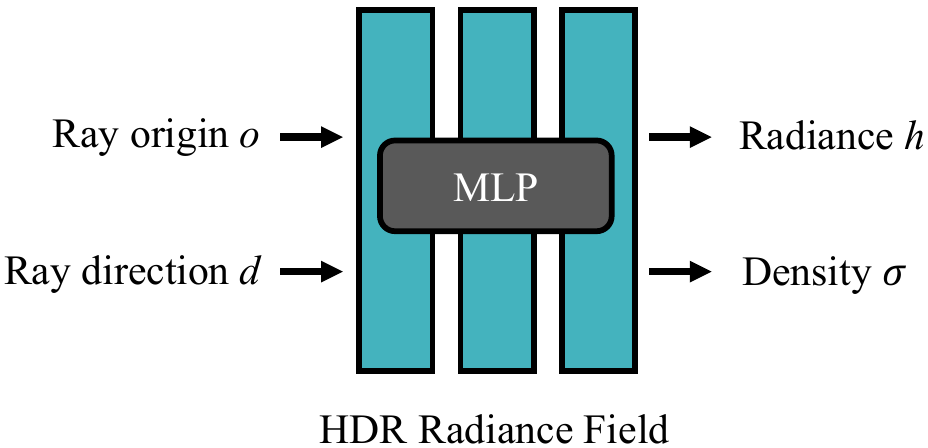}
  \caption{The inputs and outputs of the MLP implicitly model the  HDR radiance field. Image adapted from~\cite{huang2022HDRnerf}.}
  \label{fig:nerfNet}
  \vspace{-0.8em}
\end{figure}

\begin{table*}[!t]
  \centering
    \caption{Quantitative comparison of HDR novel view synthesis methods averaged across eight synthetic scenes.} \label{tab:nerf_all}
    \resizebox{\linewidth}{!}{
    \begin{tabular}{lccccccccc}
        \toprule
        \multirow{2}{*}[-3pt]{Method} &\multirow{2}{*}[-3pt]{HDR-VDP-3-$Q$}&\multirow{2}{*}[-3pt]{PSNR}&\multirow{2}{*}[-3pt]{SSIM}& \multicolumn{2}{c}{Without CRF correction}&\multicolumn{2}{c}{With CRF correction} &\multicolumn{2}{c}{$Q^\star$ (Our metric)}\\ 
        
        \cmidrule(lr){5-6} \cmidrule(lr){7-8} \cmidrule(lr){9-10} & & & & \textrm{PU21-PSNR} & \textrm{PU21-SSIM} & $\textrm{PU21-PSNR}$& \textrm{PU21-SSIM} & $Q^\star_\textrm{PSNR}$& $Q^\star_\textrm{SSIM}$ \\ 
        \midrule
        HDR-NeRF & 7.023 & 25.513 & 0.863 & 31.388 & 0.901 & 38.485 & 0.929 & 28.957 & 0.899  \\
        HDR-NeRF+ & 9.634 & 27.358 & 0.929 & 35.758 & 0.953 & 41.350 & 0.957 & 32.457 & 0.937 \\
        HDR-NeRF$\dagger$ & \textbf{9.863} & \textbf{28.754} & \textbf{0.933} & \textbf{38.202} & \textbf{0.967} & \textbf{43.483} & \textbf{0.973} & \textbf{34.539} & \textbf{0.968}  \\
        \bottomrule
    \end{tabular}}
    \vspace{-0.8em}
\end{table*}

\begin{figure*}[!t]
\centering
  \includegraphics[width=\textwidth]{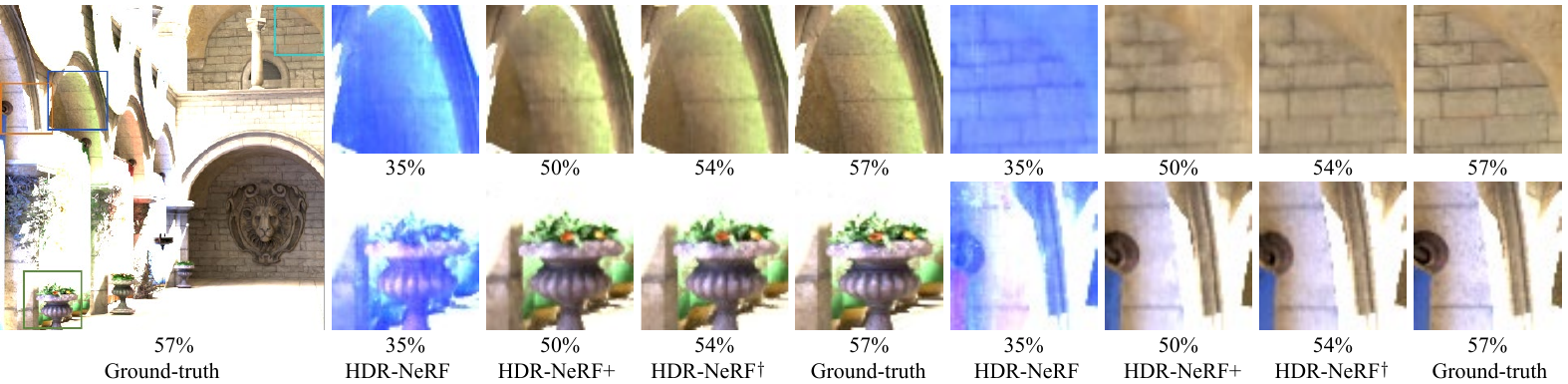}
  \caption{Visual comparison of HDR novel view synthesis methods on the ``Sponza" scene. For the reference HDR view as the ground-truth, we set the exposure value $v$ in Eq.~\eqref{eq:IDM} to be the $57$-th of the full dynamic range. Other percentages are the optimally matched $\hat{v}$ for different synthesis methods by Eq.~\eqref{eq:omeh}.}
  \vspace{-0.8em}
  \label{fig:sponza}
\end{figure*}

\subsection{HDR Novel View Synthesis}
We select HDR-NeRF~\cite{huang2022HDRnerf} as the starting point. The original HDR-NeRF employs a multilayer perceptron (MLP) to implicitly represent the radiance field of an HDR scene, and uses a separate MLP to function as a tone mapper to reconstruct multiple input LDR images of different exposures. Here, we simplify HDR-NeRF by stripping off the tone mapper, and directly reconstruct the HDR scene, guided by the proposed HDR quality metrics (see Fig.~\ref{fig:nerfNet}). We refer to the simplified method as HDR-NeRF$\dagger$.

\noindent\textbf{Network Design}.
We employ an eight-layer MLP with $256$ channels to implicitly reconstruct the HDR scene radiance. 
For a given ray $r = o + sd$, where $o$ is the origin, $d$ is the ray direction, and $s$ denotes a position along the ray, the MLP outputs the radiance $h$ and density $\sigma$, based on which the luminance value can be computed by
\begin{equation}
\label{eqn:render}
\hat{H}(r)= \int_{s_n}^{s_f} T(s)\sigma(r(s))h(r(s))ds,
\end{equation}
where
\begin{equation}
\label{eqn:trans}
T(s)= \exp\left(- \int_{s_n}^{s}\sigma(r(v))dv\right).
\end{equation}
 $s_n$ and $s_f$ denote the near and far boundary of the ray, respectively. $T(s)$ denotes the accumulated transmittance along the ray from $s_n$ to $s$.

\noindent\textbf{Loss Function}.
For computational convenience, we adopt the proposed HDR quality metric, $Q^\star_\textrm{MAE}$ (rather than $Q^\star_\textrm{DISTS}$), as the loss function to encourage high-fidelity novel view synthesis across all luminance levels. 

\subsection{Experimental Setups}
\noindent\textbf{Model Training and Testing}.
We employ the dataset in~\cite{huang2022HDRnerf}, comprising $8$ synthetic scenes rendered by Blender\footnote{https://www.blender.org/}. There are $35$ HDR views for each scene, and we select $18$ views for training, and leave the remaining $17$ for testing. The resolution of each view is $400\times 400$.

Training follows the original paper~\cite{huang2022HDRnerf}. We employ the positional encoding in~\cite{Mildenhall2021NeRF}, and optimize a coarse model and a fine model, where the density predicted by the coarse model is used to bias the sampling of a ray in the fine model.
We sample $64$ points along each ray in the coarse model and $128$ points in the fine model. 
We employ the Adam optimizer~\cite{Kingma2014adam} with an initial learning rate $5 \times 10^{-4}$, which decays exponentially to $5 \times 10^{-5}$ with a total of $200,000$ iterations. The batch size of rays is set to $1,024$.

\begin{table*}[t]
  \centering
    \caption{Quantitative comparison of HDR-NeRF$\dagger$ optimized by different loss functions.} \label{tab:nerf}
 \scalebox{0.85}{
    \begin{tabular}{lccccccccc}
        \toprule
        \multirow{2}{*}[-3pt]{Loss} &\multirow{2}{*}[-3pt]{HDR-VDP-3-$Q$}&\multirow{2}{*}[-3pt]{PSNR}&\multirow{2}{*}[-3pt]{SSIM}& \multicolumn{2}{c}{Without CRF correction}&\multicolumn{2}{c}{With CRF correction} &\multicolumn{2}{c}{$Q^\star$ (Our metric)}\\ 
        \cmidrule(lr){5-6} \cmidrule(lr){7-8} \cmidrule(lr){9-10} & & & & \textrm{PU21-PSNR} & \textrm{PU21-SSIM} & $\textrm{PU21-PSNR}$& \textrm{PU21-SSIM} & $Q^\star_\textrm{PSNR}$& $Q^\star_\textrm{SSIM}$ \\ 
        \midrule
        MAE & 6.224 & 17.181 & 0.454 & 25.842 & 0.613 & 31.577 & 0.856 & 22.928 & 0.849  \\
        PU21-MAE & 9.717 & 28.670 & 0.929 & 37.293 & 0.960 & 42.245 & 0.970 & 33.477 & 0.949 \\
        $\log$-MAE & 9.720 & 23.173 & 0.840 & 34.006 & 0.936 & 41.267 & 0.961 & 32.423 & 0.944  \\
        $\mu$-MAE & 9.784 & 25.645 & 0.894 & 35.408 & 0.948 & 42.104 & 0.968 & 33.227 & 0.948 \\
        \midrule 
        $Q^\star_\textrm{MAE}$ & \textbf{9.863} & \textbf{28.754} & \textbf{0.933} & \textbf{38.202} & \textbf{0.967} & \textbf{43.483} & \textbf{0.973} & \textbf{34.539} & \textbf{0.968} \\ 
        \bottomrule
    \end{tabular}}
    \vspace{-0.8em}
\end{table*}

\begin{figure}[!t]
\centering
  \includegraphics[width=0.38\textwidth]{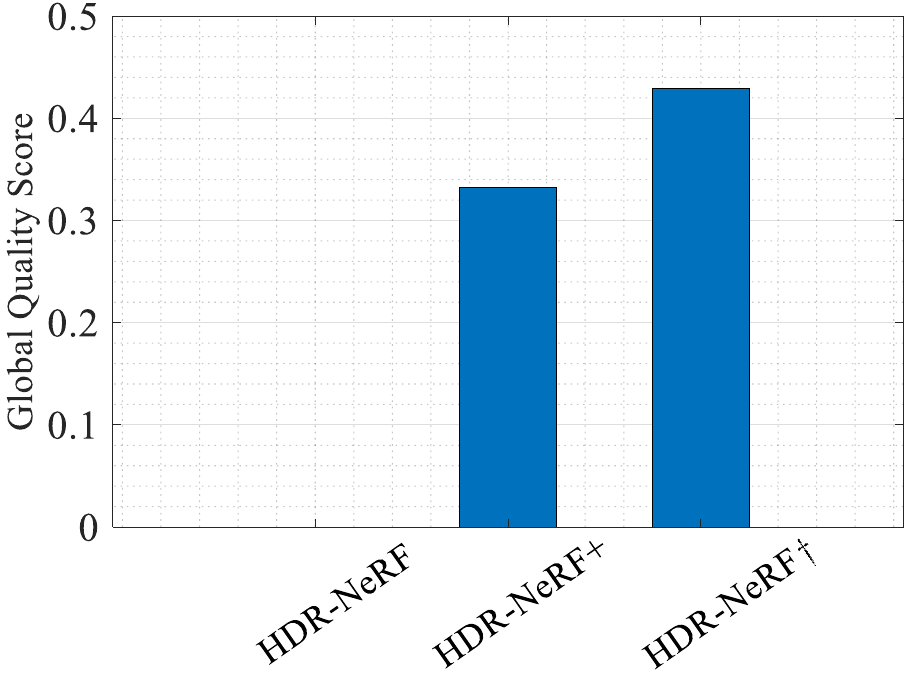}
  \caption{Subjective quality scores across all test views and observers in the 2AFC subjective user study. HDR-NeRF serves as the baseline with the global quality score of zero.}
  \label{fig:subjective_nerf}
  \vspace{-0.8em}
\end{figure}

\noindent\textbf{Competing Methods}. We compare our method against HDR-NeRF~\cite{huang2022HDRnerf} and its variant optimized for HDR views directly, denoted by HDR-NeRF+. When training  HDR-NeRF+, the reference and predicted HDR luminance values are tone mapped to LDR values by a simple TMO~\cite{Reinhard2002photographic}, as suggested in~\cite{huang2022HDRnerf}. 

\noindent\textbf{Evaluation Criteria}. We employ several objective quality metrics: 1) HDR-VDP-3-$Q$~\cite{mantiuk2023hdr} 2) PSNR, 3) SSIM~\cite{wang2004image}, 4)  PU21-PSNR, 5) PU21-SSIM, 6) PU21-PSNR with camera response function (CRF) correction~\cite{Hanji2022}, 7) PU21-SSIM with CRF correction, 8) the proposed quality metric with PSNR as the base model (\ie, $Q^\star_\textrm{PSNR}$), and 9) the proposed quality metric with SSIM as the base model (\ie, $Q^\star_\textrm{SSIM}$). The CRF correction compensates for the metric sensitivity to the shifts in tone and color~\cite{eilertsen2021cheat}, and is applied before the PU21 encoding.

\subsection{Experimental Results}
\noindent\textbf{Quantitative Evaluation}.
Table~\ref{tab:nerf_all} lists the average results of rendered novel HDR views of the eight synthetic scenes. 
The primary observation is that the proposed HDR-NeRF$\dagger$ outperforms HDR-NeRF+ by a clear margin under all evaluation metrics. This demonstrates the superiority of the adopted inverse display model over the simple tone mapper in HDR-NeRF+. 
Lack of direct supervision, HDR-NeRF performs marginally in synthesizing HDR views, despite its ability to reconstruct a satisfying output LDR image.

\noindent\textbf{Qualitative Evaluation}.
Fig.~\ref{fig:sponza} visually compares the results on a test view of the ``Sponza" scene. The HDR view synthesized by HDR-NeRF suffers from color cast and detail loss. HDR-NeRF+ recovers more details, but not as sharp as those rendered by the proposed HDR-NeRF$\dagger$.

\noindent\textbf{Subjective User Study}.
We perform a subjective user study to verify the perceptual advantages of HDR-NeRF$\dagger$. For each of the eight scenes, we randomly choose five test views,  reconstructed by the three competing methods (including HDR-NeRF$\dagger$). We manually select $\{v^{(k)}\}_{k=1}^3$ to zoom in the low, middle, and high luminance range of each reference HDR view, respectively. All LDR images are aligned to the reference LDR images by solving Eq.~\eqref{eq:omeh}. We adopt the two-alternative forced choice (2AFC) approach to gather human preferences of  $\binom{3}{2}\times 8 \times 5 \times 3 = 360$ image pairs from $15$ participants. They are given unlimited time to review the images, and are allowed to take a break at any time during subjective testing to mitigate fatigue effects. 
The global quality scores are aggregated by the maximum likelihood estimation~\cite{tsukida2011analyze}. 
Fig.~\ref{fig:subjective_nerf} shows the results, which verify the perceptual gains of HDR-NeRF$\dagger$ driven by the proposed HDR quality metric.

\noindent\textbf{Ablation study}. 
We evaluate the view synthesis performance of HDR-NeRF$\dagger$ optimized by several different quality metrics as the loss functions: 1) MAE, 2) PU21-MAE, 3) log-encoded MAE (\ie, $\log$-MAE), 4) MAE computed in the LDR domain tone mapped by the $\mu$-law~\cite{1957SmithInstantaneous} (\ie, $\mu$-MAE), and 5) the proposed $Q^\star_\textrm{MAE}$.
Table~\ref{tab:nerf} presents the quantitative comparison results, where HDR-NeRF$\dagger$ optimized by $Q^\star_\textrm{MAE}$ 
delivers the best results. The encoding-based metrics like  PU21-MAE and $\log$-MAE do not necessarily surpass $\mu$-MAE, even though tone mapping would cause detail loss and color distortion.

\section{Conclusion and Discussion}
We have described a family of HDR quality metrics by augmenting current LDR quality metrics with a simple inverse display model. Our metrics are efficient in inheriting the benefits of advanced LDR quality metrics, flexible to zoom in and align specific luminance ranges for more detailed assessment, and training-free. We have validated our HDR quality metrics in terms of correlation with human perceptual scores on four HDR-IQA datasets and perceptual optimization of HDR novel view synthesis.

Previous studies of HDR image processing are inclined to adopt a global tone mapper 
for visualizing and comparing the processed results. 
In contrast, this paper suggests an alternative visualization method of using the inverse display model in Eq.~\eqref{eq:IDM}. This method allows us to focus on and enhance the visibility of different portions of luminance ranges for a more fine-grained visual comparison (see Fig.~\ref{fig:sponza}). Together with the proposed family of HDR quality metrics, we expect more rapid and reliable progress of HDR imaging and rendering in the near future. 

As one of the limitations, our metrics do not account for the reduced sensitivity of the visual system at low luminances. That is, the predictions are the same regardless of whether the image is meant to be shown on a dark or bright display. The PU/PU21 encoding and HDR-VDP are designed to model the changes in sensitivity with absolute luminance levels.

\section*{Acknowledgments}
 This work was supported in part by the National Natural Science Foundation of China under Grant 62071407 and the Hong Kong RGC Early Career Scheme (2121382). 

{
    \small
    \bibliographystyle{ieeenat_fullname}
    \bibliography{main}
}

\end{document}